\documentclass[article,twocolumn,showpacs,floatfix,longbibliography]{revtex4-2}
\usepackage{graphicx} 
\usepackage[caption=false]{subfig}
\usepackage{hyperref}
\usepackage{xcolor}
\usepackage{multirow}
\usepackage{amsmath}

\begin{document}

\title{
Towards a Robust Model-Independent Test of the DAMA/LIBRA Dark Matter Signal: ANAIS--112 Results with Six Years of Data
}


\author{Julio Amar{\'e}\textsuperscript{1,2}}
\author{Jaime Apilluelo\textsuperscript{1,2}}
\author{Susana Cebri{\'an}\textsuperscript{1,2}}
\author{David Cintas\textsuperscript{1,2}}
\author{Iv{\'a}n Coarasa\textsuperscript{1,2}}
\email{icoarasa@unizar.es}
\author{Eduardo Garc\'{\i}a\textsuperscript{1,2}}
\author{Mar\'{\i}a Mart\'{\i}nez\textsuperscript{1,2}}
\author{Ysrael Ortigoza\textsuperscript{1,2,3}}
\author{Alfonso Ortiz~de~Sol{\'o}rzano\textsuperscript{1,2}}
\author{Tamara Pardo\textsuperscript{1,2}}
\author{Jorge Puimed{\'o}n\textsuperscript{1,2}}
\author{Mar\'{\i}a Luisa Sarsa\textsuperscript{1,2}}
\email{mlsarsa@unizar.es}
\author{Carmen Seoane\textsuperscript{1,2}}
\affiliation{\textsuperscript{1}Centro de Astropart\'{\i}culas y F\'{\i}sica de Altas Energ\'{\i}as (CAPA), Universidad de Zaragoza, Pedro Cerbuna 12, 50009 Zaragoza, Spain}
\affiliation{\textsuperscript{2}Laboratorio Subterr\'aneo de Canfranc, Paseo de los Ayerbe s.n., 22880 Canfranc Estaci\'on, Huesca, Spain}
\affiliation{\textsuperscript{3}Escuela Universitaria Polit\'ecnica de La Almunia de Do\~{n}a Godina (EUPLA), Universidad de Zaragoza, Calle Mayor 5, La Almunia de Do\~{n}a Godina, 50100 Zaragoza, Spain}
\date{\today}

\newcommand{\DL}{DAMA\slash LIBRA\ }
\newcommand{\ANAIS}{\mbox{ANAIS--112}\ }
\newcommand{\COSINE}{COSINE--100\ }

\newcommand{\ckkd}{c/keV/kg/d}
\newcommand{\keV}{keV$_{\textnormal{ee}}$}
\newcommand{\keVnr}{keV$_{\textnormal{nr}}$}

\newcommand{\tritium}{$^{3}$H\ }
\newcommand{\Na}{$^{22}$Na\ }
\newcommand{\K}{$^{40}$K\ }
\newcommand{\NaK}{$^{22}$Na\slash$^{40}$K\ }
\newcommand{\Cd}{$^{109}$Cd\ }
\newcommand{\Cf}{$^{252}$Cf\ }
\newcommand{\I}{$^{127}$I\ }
\newcommand{\Pb}{$^{210}$Pb\ }

\newcommand{\Sen}{\mathcal{S}}

\renewcommand\labelenumi{(\theenumi)}

\begin{abstract}

The nature of dark matter, which constitutes 27\% of the Universe's matter-energy content, remains one of the most challenging open questions in physics. Over the past two decades,
the DAMA/LIBRA experiment has reported an annual modulation
in the detection rate of $\approx$250~kg of NaI(Tl) detectors operated at the Gran Sasso Laboratory, which the collaboration interprets as evidence of the galactic dark matter detection.  
However, this claim has not been independently confirmed and is refuted under certain dark matter particle and halo model scenarios. Therefore, it is crucial to perform an experiment with the same target material. The ANAIS experiment uses 112.5 kg of NaI(Tl) detectors at the Canfranc Underground Laboratory and it has been collecting data since August 2017 to model-independently test the DAMA/LIBRA result. This article presents the results of the annual modulation analysis corresponding to six years of ANAIS--112 data. Our results, the most sensitive to date with the same target material, NaI(Tl), are incompatible with the DAMA/LIBRA modulation signal at a 4\,$\sigma$ confidence level. Such a discrepancy strongly challenges the DAMA/LIBRA dark matter interpretation and highlights the need to address systematic uncertainties affecting the comparison, particularly those related to the response of detectors to nuclear recoils, which may require further characterization of the DAMA crystals.

\end{abstract}

\maketitle


While astronomical and cosmological evidence about dark matter (DM) accumulates, its nature remains elusive~\cite{ParticleDataGroup:2024cfk}. 
Over the past thirty years, various excesses above background have been reported in both indirect and direct dark matter searches, most of which have been resolved.
However, a few of them persist and deserve further investigation~\cite{Leane:2022bfm}. One of the most puzzling results is that from
the \DL experiment, taking data at the Laboratori Nazionali del Gran Sasso (Italy), which reports a positive detection for more than twenty years~\cite{Bernabei:2020mon, Bernabei:2021kdo}. \DL detectors observe an annual modulation in the detection rate at low energies, below 6 keV (in electron-equivalent energy~\footnote{In this Letter, we will use keV for referring to electron-equivalent energies while nuclear recoils energies will be expressed in \keVnr}), consistent with the predictions of weakly interacting 
massive particles (WIMPs) distributed in the galactic halo~\cite{Drukier:1986tm, Freese:1987wu}, and having cross-sections with nucleons in the range of $10^{-40}-10^{-42}$~cm$^2$~\cite{Bernabei:2020mon}. 
The statistical significance of such a detection is overwhelming, but the origin of such a modulation has not yet been determined. No other experiment has observed a compatible signal~\cite{Billard:2021uyg, Cooley:2022ufh}, but the uncertainties and unknowns in both the dark matter particle and halo models make it very difficult to disprove the interpretation of the result in terms of a DM signal in a model-independent way~\cite{Leane:2022bfm}.
This is the objective of ongoing or recently decommissioned experiments, such as \ANAIS~\cite{Amare:2018sxx, Amare:2019jul, Amare:2019ncj, Amare:2021yyu, Coarasa:2024xec} and COSINE--100~\cite{COSINE-100:2019lgn, COSINE-100:2021zqh}, or other upcoming projects like COSINE-100U~\cite{Lee:2024wzd}, COSINUS~\cite{COSINUS:2023kqd}, SABRE~\cite{SABRE:2018lfp, Calaprice:2022vte, Zurowski:2023ndx}, and PICOLON \cite{picolon2021}.
\par
The \DL experiment is expected to have completed the data taking by the end of 2024. The final results should then be released.  
The \COSINE experiment was in operation from 2016 to 2023 in the YangYang Underground Laboratory in Korea. After concluding data collection, the \COSINE collaboration recently released results from the full dataset~\cite{Carlin:2024maf}, corresponding to 6.4~years of operation and a total effective exposure of 358~kg$\times$yr.
The \ANAIS experiment, in operation since August 2017 at the Canfranc Underground Laboratory (LSC)
in Spain, consists of 112.5~kg of NaI(Tl), distributed in nine detectors, labeled in the following D0 to D8, and produced by the same provider as those used in the \COSINE experiment (Alpha Spectra Inc.).
A blank module, similar in design to the other modules but without the NaI(Tl) crystal, operates in the same experimental space within an independent shielding, but integrated into the same data acquisition system, to monitor non-NaI(Tl) scintillation events.
The trigger threshold is set-up at photoelectron level in each PMT and a coincidence within 200~ns between the two PMT signals of the same module is required for triggering that module. Multiple-hit events are defined as those where two or more module-triggers occur within a coincidence window of 1~$\mu$s, while for single-hit events only one module triggers in that window. In ANAIS, the muon vetoes are readout by an independent DAQ system and they do not contribute neither to the definition of multiple-hit events nor to the dead time. 
Details on the \ANAIS experimental set-up, detector performance and previous results can be found elsewhere~\cite{Amare:2018sxx, Amare:2019jul, Amare:2019ncj,  Amare:2021yyu, Coarasa:2022zak, Coarasa:2024xec}. \ANAIS will complete more than 8 years of data by the end of 2025, achieving a sensitivity exceeding the 5\,$\sigma$ confidence level according to our prospects (see Figure~9 within the Supplemental Material~\cite{supp}).
Here, we present the analysis of the annual modulation corresponding to six years of \ANAIS data for a total effective exposure of 625.75~kg$\times$yr, which confirms our sensitivity estimates.
\par
First, we briefly describe the key features of the analysis pipeline applied to the 6 years data, which in some aspects differs from that applied to obtain previous results~\cite{Amare:2019jul, Amare:2019ncj, Amare:2021yyu, Coarasa:2024xec}.
\par
A more robust calibration procedure for the low energy range has been developed and applied. We use a proportional calibration using the $^{109}$Cd line at around 22\,keV as reference to correct any possible gain drift. Later, by combining the whole exposure, we apply a recalibration procedure specifically designed for the ROI using information from two internal contaminants in the NaI(Tl) crystals, $^{22}$Na and $^{40}$K, both homogeneously distributed in the crystal bulk. They provide distinct peaks at 0.9 and 3.2\,keV in the module where the decay occurs.
These peaks are identified through coincidences with a high-energy gamma ray from the nuclear de-excitation, which can escape and be detected in a second module.
The corresponding evolution of the rates for both peaks, shown in Figure~1 within the Supplemental Material~\cite{supp}, is fully consistent with a constant rate in the case of \K (T$_{1/2}$(\K)=1.25 10$^9$\,yr) and shows an exponential decay compatible with the isotope lifetime in the case of \Na (1389$\pm$51~d is the lifetime derived from the fit, while 1369~d is the nominal lifetime for $^{22}$Na). This evolution of the \Na low-energy events  
(just below the ANAIS--112 analysis threshold, set at 1\,keV) along the 6 years analyzed, guarantees both the stability of the threshold and the accuracy of the calibration in the ROI.
\par

\begin{figure}[!htbp]
    \centering
    \includegraphics[width=0.48\textwidth]{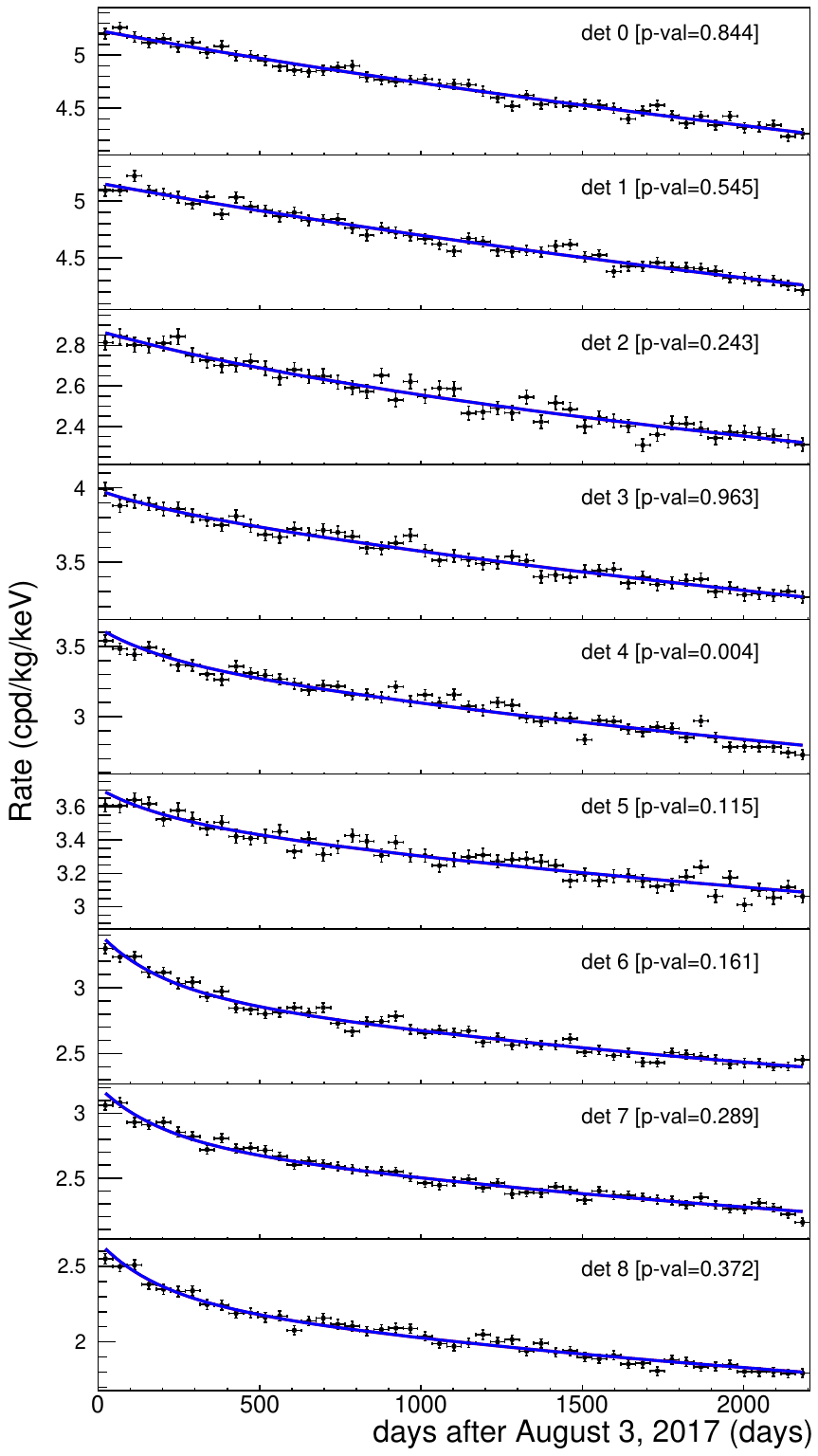}
    \caption{Results of the fit for the data from the nine modules in the [1--6]\,keV energy region, under the modulation (blue) and null hypotheses (red). In all the panels, the red line is masked by the blue one, as the fit obtained for the modulated hypothesis is consistent with $S_m=0$. p-values under the modulation hypothesis are individually displayed for each module. The global results of the fit are: for the null hypothesis, $\chi^2$/ndf~=~451.34/423, (p-value~=~0.164), and for the modulation hypothesis, $\chi^2$/ndf~=~451.31/422, (p-value~=~0.156). The best-fit modulation amplitude in the latter case is $S_m = (-0.4 \pm 2.5)$ cpd/ton/keV.}
    \label{fig:rateEvol16}
\end{figure}

A new filtering procedure to remove non-bulk scintillation events is applied, following Ref.~\cite{Coarasa:2022zak, Coarasa:2023lhj}, as done in Ref.~\cite{Coarasa:2024xec} for the reanalysis of the 3-year exposure. 
This filtering protocol is based on Boosted Decision Trees (BDT) and the training does not use background events at all, but uses events from $^{252}$Cf calibrations in the [1--2]\,keV energy region resulting mainly from nuclear elastic scattering from neutrons as signal and non-NaI(Tl) bulk scintillation events from the blank module in an equivalent energy range, as noise. 
The efficiency of this filtering procedure is derived both from $^{252}$Cf and $^{109}$Cd calibrations data, and their respective time evolution in the [1--6]\,keV energy region along the data taking is shown in the Figure~3 of the Supplemental Material~\cite{supp}.
By considering the deviations of the \Cd efficiency averaged for all the detectors with respect to their mean value in the [1--8]\,keV energy region, we obtain a standard deviation of 0.13\,$\%$, while it increases to 0.3\,$\%$ by considering the detectors independently. These values are better or of the same order that those reported for DAMA/LIBRA-phase2 efficiencies in that energy region (0.3\,$\%$)~\cite{Bernabei:2020mon}. 
It can be observed ~\cite{supp} that \Cf and \Cd efficiencies are compatible. In the current analysis, the average of the efficiencies in the ROI derived from the seven \Cf calibrations will be used.
However, according to the evolution of the \Cd efficiency, constant efficiencies have been assumed for all modules except D0, D4 and D5. 
During the first year of data collection, modules D4 and D5 showed instabilities until the operation HV was reduced. Consequently, a different (but still constant) efficiency is used in the analysis for D4 during the first 361~days and for D5 during the first 153~days. In the case of D0, one of the PMTs suffered from a gain loss in the last months of the sixth year, and then, a different (but constant) efficiency is used in the analysis for the last $\approx$200~days of that year.
\par
This new filtering has allowed to significantly increase the efficiency for bulk scintillation event selection, but the remaining background below 3\,keV is still showing an excess with respect to our background modeling estimates~\cite{Amare:2018ndh, Coarasa:2022zak}. Work is in progress in complementary directions towards understanding such an excess.
\par
After selecting bulk scintillation events by the cut on the BDT parameter, the event rate is evaluated in one-day binning using the corresponding live time. 
High-rate periods which are more than 3 standard deviations over the annually averaged detection rate below 3\,keV are removed (see Ref.~\cite{Coarasa:2024xec} for more details)~\footnote{The effect of this cut on the annual modulation analysis presented later in this manuscript was checked with a toy MC. By injecting a modulation with the DAMA/LIBRA amplitude over a background sampled with ANAIS--112 features and applying a cut on periods having a rate at ±3 sigma level of the average, the bias observed in the modulation amplitude recovered after the analysis is (0.00 $\pm$ 0.02) cpd/ton/keV.}. 

The time corresponding to the rejected periods is discounted from the effective live time used for calculating the effective exposure. The events arriving within 1\,s from a cosmic muon triggering the veto system had been removed before applying the BDT cut and the live time had also been conveniently corrected~\cite{Amare:2018sxx}. 
\par

The accumulated exposure used for the annual modulation analysis corresponding to the six years of data of the \ANAIS experiment is summarized in Table~1 within the Supplemental Material~\cite{supp}. It also details the dead time, down time, percentage of live time rejected in the analysis, and the corresponding effective exposure after subtracting the latter. 
\par
Then, the rate is recalculated for each detector in the energy windows of interest with the corresponding live time and corrected by the average efficiency for each energy window in 0.1-keV bins, before applying the annual modulation analysis.
The background after applying the filtering procedure corresponding to the 6 years data for all the nine modules in the ROI is shown in Figure~4 within the Supplemental Material~\cite{supp}, compared to the corresponding background of \COSINE and \DL experiments.
\par
We search for the modulation in the overall event rate over time through a least squares fit, by defining the $\chi^2$ function as follows:
\begin{equation}
    \chi^2 = \sum_{i,d} \frac{(n_{i,d} - \mu_{i,d})^2}{\sigma_{i,d}^2} ,
\end{equation}
where $n_{i,d}$ represents the number of events in the ROI in the time bin $t_i$ for detector $d$, obtained by correcting 
the measured event count using the live time for that specific temporal bin and detector, along with the corresponding acceptance efficiency;
$\sigma_{i,d}$ is the Poisson uncertainty associated with the event count, also corrected by the corresponding live time and efficiency, whose uncertainties are combined in quadrature; and
$\mu_{i,d}$ denotes the expected number of events in that particular time bin and detector, including a hypothetical dark matter component.
\par
$\mu_{i,d}$ is expected to diminish over time because there are background contributions from radioactive isotopes with half-lives on the order of a few years, primarily $^{210}$Pb (T$_{1/2}$=22.3\,yr), $^{3}$H (T$_{1/2}$=12.3\,yr) and $^{22}$Na (T$_{1/2}$=2.6\,yr). For detectors D6, D7 and D8, contributions from Te and I, cosmogenically produced isotopes with shorter half-lives, are also relevant. 
Accurately modeling this background rate decrease is crucial to avoid biasing the fit. Our background modeling is based on the independent determination of the contamination levels of the crystals and other detector components by different techniques \footnote{Crystal and other detector components contamination levels introduced in our background modeling have been determined by using different techniques. For the most relevant crystal contaminations we used coincidence measurements (for $^{40}$K, $^{22}$Na and other cosmogenically produced isotopes), alpha rate determination (for $^{210}$Pb-$^{210}$Bi-$^{210}$Po), and Bi-Po sequences (for $^{238}$U and $^{232}$Th chains). For other detector components, HPGe spectroscopy was used, being the PMTs the most relevant background contribution. A thorough description can be found in ~\cite{Amare:2018ndh}.}, followed by the Monte Carlo (MC) simulation within the Geant4 package~\cite{Amare:2018ndh}. This background modeling is used to describe the background evolution in time for every detector.  
\par
Then, we model $\mu_{i,d}$ as:
\begin{multline}
    \mu_{i,d}=[R_{0,d}(f_d\phi_{bkg,d}^{MC}(t_i)+(1-f_d)\phi_{flat}(t_i)) \\
    +S_m \cos(\omega(t_i-t_0))]M_d\Delta E \Delta t,
    \label{eq:modFit}
\end{multline}
where $\phi_{bkg,d}^{MC}$ is the probability distribution function 
sampled from the MC model,
describing the background evolution at time bin $t_i$ for detector $d$;
$\phi_{flat}$ is a constant probability distribution function that accounts for the noise contribution not explained by the background model (related to the excess below 3\,keV previously commented) but found at a constant rate in the data, and the average component of a hypothetical contribution from DM interactions;
$M_d$ is the mass of every module; and $\Delta E$ and $\Delta t$ represent energy and time intervals, respectively.
$R_{0,d}$ and $f_d$ are nuisance parameters in the fitting procedure: 
$R_{0,d}$ represents the background index in the considered energy region, while $f_d$ measures the relative weight of the MC estimated time dependence with respect to the total rate.
$S_m$ represents the DM annual modulation amplitude, and $\omega$ and $t_0$ are the angular frequency and the phase of the modulation searched for in the data. For the standard halo model, they correspond to a period of 1 year and June, 2, respectively. 
$S_m$ is set to 0 to test the null hypothesis and allowed to vary freely for the modulation hypothesis.

\begin{table*}[htbp!]
\centering
\begin{tabular*}{0.85\textwidth}{@{\extracolsep{\fill}}ccccc}
\hline\hline 
Energy region & Experiment & Exposure (kg$\times$y) & p-value mod hyp & $S_m$ (cpd/ton/keV) \\

\hline
\multirow{3}{*}{\shortstack{[1--6] keV}}
    & ANAIS    &  625.75 & 0.156 & -0.4$\pm$2.5 \\
    & COSINE &  358.00 &  $\cdot\cdot\cdot$  &  1.7$\pm$2.9 \\    
    & DAMA       & 1126.40 & 0.513 & 10.5$\pm$1.1 \\
\hline
\multirow{3}{*}{\shortstack{[2--6] keV}}
    & ANAIS    &  625.75 & 0.596 &  1.1$\pm$2.5 \\
    & COSINE &  358.00 &  $\cdot\cdot\cdot$  &  5.3$\pm$3.1 \\    
    & DAMA     & 2462.09 & 0.935 & 10.2$\pm$0.8 \\
\hline
              &            &               &         & (cpd/ton/3.3~\keVnr) \\
\hline 
\multirow{3}{*}{\shortstack{[6.7--20] \keVnr}}
    & ANAIS   &  625.75 & 0.566 &  0.0$\pm$2.3 \\
    & COSINE &  358.00 &  $\cdot\cdot\cdot$  &  1.3$\pm$2.7 \\ 
    & DAMA     & 2462.09 & 0.935 & 10.2$\pm$0.8 \\
\hline\hline
\end{tabular*}
\caption{Summary of the fit results (goodness of the fit and best fit value for the modulation amplitude) searching for an annual modulation with fixed phase in [1--6] and [2--6]\,keV energy regions for \ANAIS six-years data (this work), \COSINE full dataset~\cite{Carlin:2024maf},
and DAMA/LIBRA~\cite{Bernabei:2020mon}. Results in the [6.7--20]~\keVnr{} sodium nuclear recoil energy region  (corresponding to [2--6]\,keV for DAMA/LIBRA) are also shown.}
\label{tab:modResults}
\end{table*}

%
\begin{table*}[htbp!]
\centering
\begin{tabular*}{0.85\textwidth}{@{\extracolsep{\fill}}ccccc}
\hline\hline
Energy region    & Bias [null hyp] & $\sigma(S_m)$ [null hyp] & Bias [DAMA $S_m$]  & $\sigma(S_m)$  [DAMA $S_m$] \\
                 & (cpd/ton/keV)   & (cpd/ton/keV) & (cpd/ton/keV)      & (cpd/ton/keV)  \\
\hline 
[1--6]\,keV       & 0.01$\pm$0.02 & 2.34$\pm$0.01 & 0.00$\pm$0.02 & 2.32$\pm$0.01 \\ \relax
[2--6]\,keV       & 0.00$\pm$0.02 & 2.49$\pm$0.01 & 0.01$\pm$0.02 & 2.50$\pm$0.01 \\
\hline
                 & (cpd/ton/3.3~\keVnr) & (cpd/ton/3.3~\keVnr) & (cpd/ton/3.3~\keVnr) & (cpd/ton/3.3~\keVnr) \\
\hline
[6.7--20]~\keVnr & 0.00$\pm$0.02 & 2.23$\pm$0.01 & 0.01$\pm$0.02 & 2.25$\pm$0.01 \\
 
\hline \hline
\end{tabular*} 
\caption{Bias (true value - fitted value) of the fitting procedure derived from 20\,000 MC simulations assuming no modulation present (second column) and \DL observed modulation (fourth column). The third and fifth columns are the standard deviations of the distribution of the best fit modulation amplitudes obtained from the MC for both hypotheses.}
\label{tab:bias}
\end{table*}

\par
In the fit, the period and the phase are fixed at one year and to June 2, respectively, in order to directly compare ANAIS--112 with \DL 
results, as they appear in Ref.~\cite{Bernabei:2020mon}.
We perform two independent fits: in the energy region [2--6]\,keV, 
which can be compared with the results from the total accumulated exposure of DAMA/NaI and DAMA\slash LIBRA, and in the [1--6] keV region, which can be compared with those of DAMA/LIBRA-phase2. 
Results of the fit in the [1--6]\,keV energy region are shown in Figure~\ref{fig:rateEvol16}.
Table~\ref{tab:modResults} shows the results of the fit in the different energy regions analyzed for \ANAIS six-years data (this work), together with those of \COSINE full dataset~\cite{Carlin:2024maf}, 
and DAMA/LIBRA~\cite{Bernabei:2020mon}. It can be observed that the \ANAIS results for 6-year exposure are compatible with the absence of modulation within one standard deviation and incompatible with \DL 
at 4.0 and 3.5\,$\sigma$ C.L. for [1--6] and [2--6]\,keV energy regions, respectively~\footnote{These values take into account the uncertainty in the \DL result, unlike our previous publications. The improvement in sensitivity of our result makes more relevant the contribution from the \DL uncertainty.}. 
More information is available in the Supplemental Material~\cite{supp}: fit results corresponding to the [1--6]\,keV, [2--6]\,keV energy regions with the individual p-values and $\chi^2$/ndf (Figures~5 and 6),  residuals after subtracting the non-modulated component for each detector with the \DL modulation signal superimposed (Figure~8), and a table with the results obtained for all the nuisance parameters considered in the fits, Table~2).
\par

We assess our sensitivity to the \DL signal as the ratio $S_m^{\text{DAMA}}/\sigma(S_m)$, 
which directly gives in $\sigma$ units the C.L. at which we can test the \DL result. Then, the standard deviations for the modulation amplitude obtained in the best fit, 
\linebreak 
$\sigma(S_m)=2.5\,$\,cpd/ton/keV for both [1--6]\,keV and 
\linebreak \relax
[2--6]\,keV, 
correspond to sensitivities of (4.2$\pm$0.4)$\,\sigma$ and (4.1$\pm$0.3)$\,\sigma$, respectively, where the uncertainty corresponds to the 68\% C.L. \DL result uncertainty. Alternatively, if we define our sensitivity as $S_m^{\text{DAMA}}/\sigma'(S_m)$, where $\sigma'(S_m)$ is the quadrature combination of both uncertainties,
we obtain statistical significance of 3.8$\,\sigma$ and 3.9$\,\sigma$ for the [1--6] and [2--6] keV energy regions, respectively.

This is a highly significant result, showing strong inconsistency with the DAMA/LIBRA signal at an unprecedented level of statistical significance.
\par
We have studied how some systematics could affect such a direct comparison between \ANAIS and \DL results using toy MC simulations of experiments equivalent to 6~years of \ANAIS data, with and without adding the modulation observed by the \DL experiment.  
Table~\ref{tab:bias} shows the bias of the fitting procedure estimated from this analysis for both the modulation and null hypotheses, besides the standard deviation obtained from the fits. No bias is observed and similar standard deviations than found in the \ANAIS six-years results, as shown in Table~\ref{tab:modResults}, are obtained. We have also considered the effect of introducing additional fluctuation in the efficiency, as described in more detail in the End Matter section, finding negligible results in all cases.
\par

Although the inconsistency between the current \ANAIS results and \DL is indisputable in the case of DM particles releasing the energy by electron recoils (ER) in the NaI(Tl) crystal, for DM particles producing nuclear recoils (NR) there is a relevant source of systematic uncertainty affecting the comparison between both experiments. NR in NaI(Tl) result in a much lower light yield than ER releasing the same energy in the material. This is estimated by the relative scintillation efficiency factor or quenching factor (QF), the amount of light produced by a NR with respect to that produced by an ER. Sodium and iodine scintillation QF have been measured by several authors (see Ref.~\cite{Cintas:2024pdu} and references therein) and results do not fully agree. This hints either at a dependence on the particular crystal properties (impurities, defects, growth method, etc.) or to systematics in the calculation of the factors (as found in Ref.~\cite{Cintas:2024pdu}). In particular, QF values measured by DAMA/LIBRA, both for sodium and iodine NR, are higher than most of the recent measurements, which, in contrast to DAMA/LIBRA, also point to an energy dependence. 
Further work to improve the understanding and modeling of the scintillation QF is required to solve this issue, but we can analyse the effect of introducing different QF values in the comparison of the results. If we consider the values published by the \DL collaboration for their crystals (QF$_{\text{Na}}$=0.3 and QF$_{\text{I}}$=0.09~\cite{Bernabei:1996vj}) and the values obtained for crystals grown in the same batch than \ANAIS crystals assuming constant-with-the-energy QF values (QF$_{\text{Na}}$=0.210$\pm$0.003 and QF$_{\text{I}}$=0.066$\pm$0.022~\cite{Cintas:2024pdu}), it can be noticed that both fulfill a $\approx$3/2 proportionality, allowing to convert the [2--6]\,keV energy region from \DL into a nuclear recoil energy range of [6.7--20]~\keVnr{} for sodium recoils and [22.2--66.7]~\keVnr{} for iodine recoils, which can be directly compared with the corresponding region in ANAIS--112 data ([1.3--4.0]\,keV). We have repeated all of the previous fitting procedures in this energy region. Results are again consistent with the absence of modulation and incompatible with \DL at 4.2\,$\sigma$ C.L., as shown in Table~\ref{tab:modResults} and in Figure~7 within the Supplemental Material~\cite{supp}. 
\par
\begin{figure}
    \centering
    \includegraphics[width=0.5\textwidth]{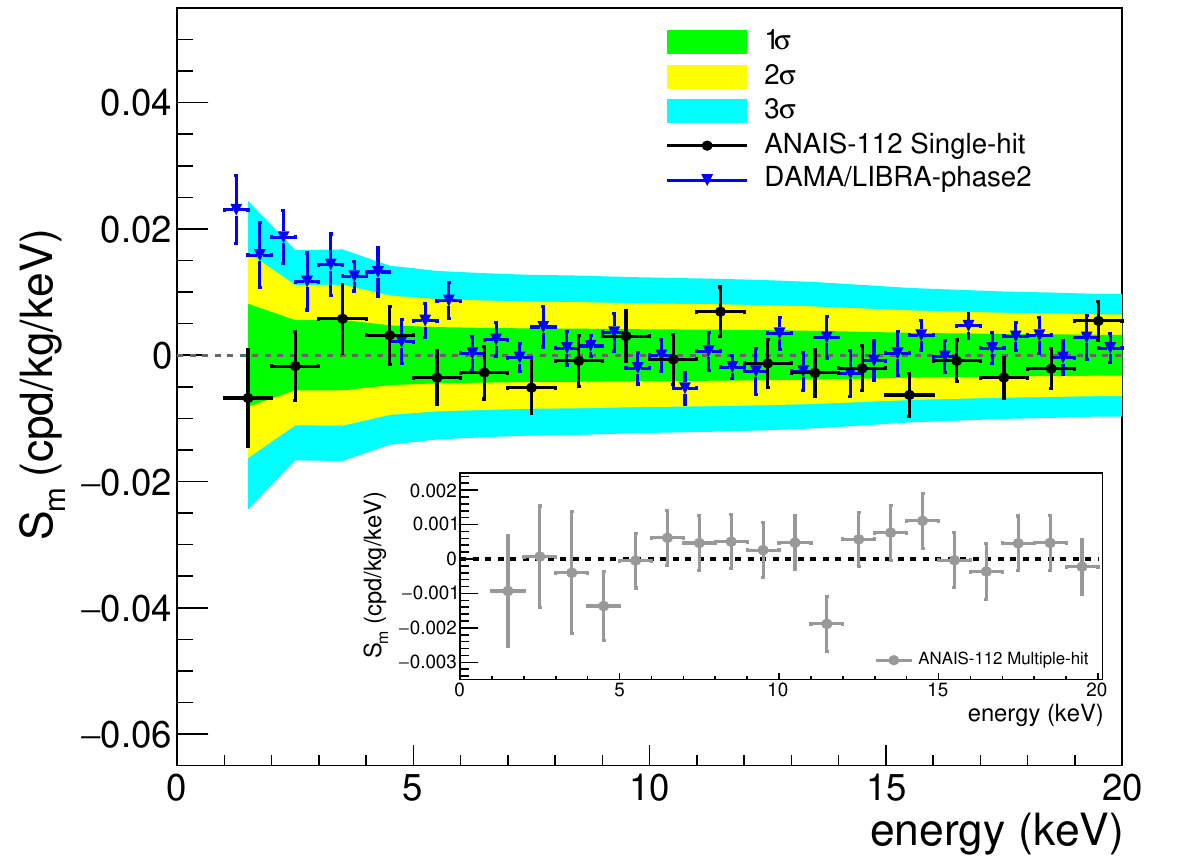}
    \caption{
     Modulation amplitude per 1\,keV energy bins for single-hit events (black dots) compared with the corresponding DAMA/LIBRA-phase2 result~\cite{Bernabei:2020mon} (blue triangles). The 1$\sigma$, 2$\sigma$ and 3$\sigma$ \ANAIS bands derived from sensitivity estimates are also shown. Inset: The same for multiple-hit events.}
     \label{fig:SmEee}
\end{figure}

Although the impact of systematics related to the uncertainty in the
QF for NaI(Tl) in the ROI for the DM analysis has not been fully estimated, 
analyzing the dependence of the modulation amplitude with the energy derived from the different experiments could help to confirm or rule out the presence of a modulation. We have analyzed the dependence of the modulation amplitude in 1-keV energy bins, from 1 to 20 keV, both for single-hit and multiple-hit events, combining the data from the nine modules. Results are shown in Figure~\ref{fig:SmEee} besides those of DAMA/LIBRA-phase2~\cite{Bernabei:2020mon}, as well as  1\,$\sigma$, 2\,$\sigma$ and 3\,$\sigma$ bands derived from sensitivity estimates from \ANAIS data~\cite{Coarasa:2018qzs}. The ANAIS results are compatible with the absence of modulation in all of the energy bins.
\par

%
%
According to the distribution of the modulation amplitude values, 
$\chi^2$/ndf~=~24.03/5 (p-value~=~2.14$\times10^{-4}$) is obtained for the \DL hypothesis, taking into account \DL result uncertainty, and $\chi^2$/ndf~=~3.14/5 (p-value~=~0.678) for the null hypothesis, both corresponding to single-hit events in the \linebreak \relax [1--6]\,keV energy region. Similar values are obtained in the [2--6]\,keV energy region.
For events with multiplicity 2, we obtain $\chi^2$/ndf~=~2.26/5 (p-value~=~0.812) in the [1--6]\,keV energy region, in full agreement with the results for single-hits.
\par
\begin{figure}
    \centering
    \includegraphics[width=0.5\textwidth]{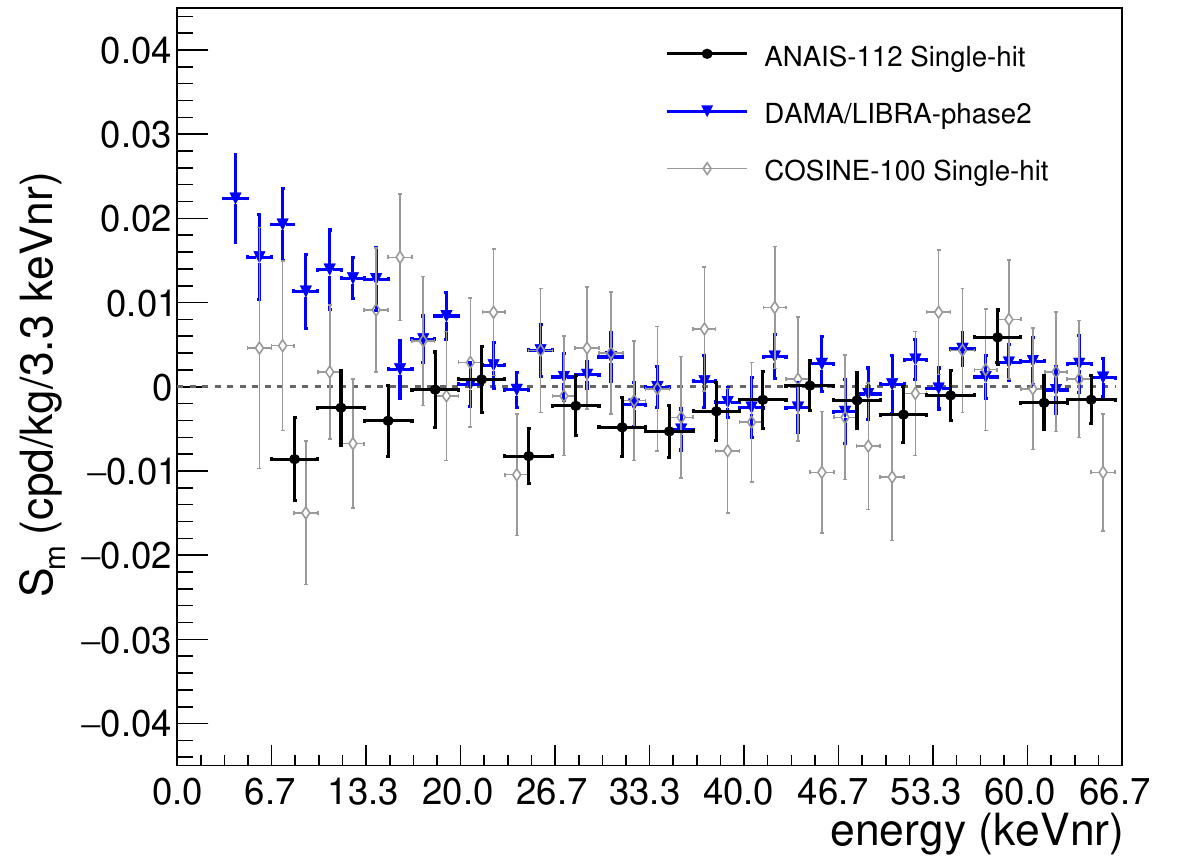}
    \caption{Comparison of the modulation amplitude for single-hit events in the energy scale of  sodium nuclear recoils from \ANAIS (black dots) and COSINE--100 data~\cite{Carlin:2024maf} (gray diamonds). The DAMA/LIBRA-phase2 result~\cite{Bernabei:2020mon} is also shown as blue triangles.}
     \label{fig:SmEnr}
\end{figure}

Figure~\ref{fig:SmEnr} shows the modulation amplitude on the scale of sodium nuclear recoils, for 3.3 \keVnr{} energy bins, for single-hit events, combining the data from the nine ANAIS modules (black dots). DAMA/LIBRA-phase2 (blue triangles) and \COSINE data (gray diamonds) are also shown for comparison. This figure assumes constant QF for sodium: 0.2 in the case of \ANAIS detectors 
and 0.3 in the case of DAMA/LIBRA. Because the relation between QF for sodium and iodine in \DL and \ANAIS is the same, the energy scale of this figure can be directly converted into iodine nuclear recoils for both datasets, but not for COSINE--100 data. The ANAIS modulation amplitudes derived within this assumption in the [6.7--20]~\keVnr{} correspond to $\chi^2$/ndf~=~4.22/4 (p-value~=~0.376) for the null hypothesis, and $\chi^2$/ndf~=~22.98/4 (p-value~=~1.28$\times10^{-4}$) for the \DL modulation hypothesis, supporting again with high statistical significance the incompatibility between both experimental results. It is worth highlighting that Figure~\ref{fig:SmEnr} is not a simple re-scaling of the energy axis from Figure~\ref{fig:SmEee}.
\par
\par
Summarizing, 6-year exposure \ANAIS results are incompatible with \DL at 4.0 and 3.5\,$\sigma$ C.L. for [1--6] and [2--6]\,keV energy regions and at 4.2\,$\sigma$ C.L. in the [6.7--20] ([22.2--66.7])\,\keVnr{} region assuming a constant QF for sodium (iodine), while compatible with full dataset COSINE-100 results, as shown in Figure~\ref{fig:comp}. According to our sensitivity prospects (see Figure~9 within the Supplemental Material), the scheduled \ANAIS full dataset (completed by the end of 2025) will be able to provide a robust refutation of the \DL signal at 5\,$\sigma$ level for ER and for NR, even assuming different scintillation QF for \DL and \ANAIS crystals~\cite{Cintas:2024pdu, Bernabei:1996vj}. 
However, it is worth to remark that the comparison in terms of NR is affected by relevant systematics related with the uncertainties in the knowledge of the scintillation quenching factors for sodium and iodine recoils. The non-proportionality of the NaI(Tl) light yield in the ROI relevant for the annual modulation analysis may partially contribute to these uncertainties. Therefore, a better modeling of both, the quenching factors and the non-proportionality of NaI(Tl) is needed. We are working to improve the estimates of the QF in \ANAIS crystals using onsite \Cf calibrations, but a better evaluation of the QF for \DL crystals is also necessary.

\section*{Acknowledgements}
This work has been financially supported by MCIN/AEI/10.13039/501100011033 under grant PID2022-138357NB-C21 and PID2019-104374GB-I00, the Consolider-Ingenio 2010 Programme under grants MultiDark CSD2009-00064 and CPAN CSD2007-00042, the LSC Consortium, the Gobierno de Arag\'on and the European Social Fund (Group in Nuclear and Astroparticle Physics) and funds from European Union NextGenerationEU/PRTR (Planes complementarios, Programa de Astrof\'{\i}sica y F\'{\i}sica de Altas Energ\'{\i}as). Authors would like to acknowledge the use of the Servicio General de Apoyo a la Investigaci\'on-SAI, Universidad de Zaragoza and technical support from LSC and GIFNA staff.
\\

\section*{Author contributions statement}

The main analysis presented in this article was conducted by IC supervised by MM and MLS. The manuscript was written by IC, MM and MLS. All authors have read and agreed to the published version, and are listed alphabetically by their last names. 
EG, AOdS and MLS contributed to the experiment design; JA, SC, AOdS and MLS contributed to the experiment setting-up and commissioning; JA, JAp, SC, DC, IC, MM, YO, AOdS, TP, CS and MLS performed calibration and maintenance tasks; MM developed DAQ software tools; IC, MM and MLS developed analysis tools; IC, EG, MM, JP and MLS analyzed the data; SC and TP developed background simulations; EG, AOdS and JP performed radiopurity measurements; IC, EG, MM and JP contributed to sensitivity estimates.

\section*{Data and code availability}
The data that support the findings of this Letter are openly available at the website of the ORIGINS Excellence Cluster: \url{https://www.origins-cluster.de/odsl/dark-matter-data-center/available-datasets/anais}.

\section*{References}
\bibliographystyle{apsrev4-2}
\bibliography{biblio}


\section*{End Matter}

\subsection*{Analysis of possible bias associated to variations in the efficiency with toy MC}

We have carried out 20\,000 toy MC simulations of experiments equivalent to 6~years of \ANAIS data (taking into account both the time-dependent background evolution and the measured efficiencies), with and without adding the modulation observed by the \DL experiment. Table~\ref{tab:bias} shows the bias of the fitting procedure estimated from this analysis for both the modulation and null hypotheses, besides the standard deviation obtained from the fits. No bias is observed and similar standard deviations than found in the \ANAIS six-years results, as shown in Table~\ref{tab:modResults}, are obtained.
\par
\begin{figure*}
    \begin{center}
    \begin{minipage}[b]{0.3\linewidth}
        \centering
        \includegraphics[width=\linewidth]{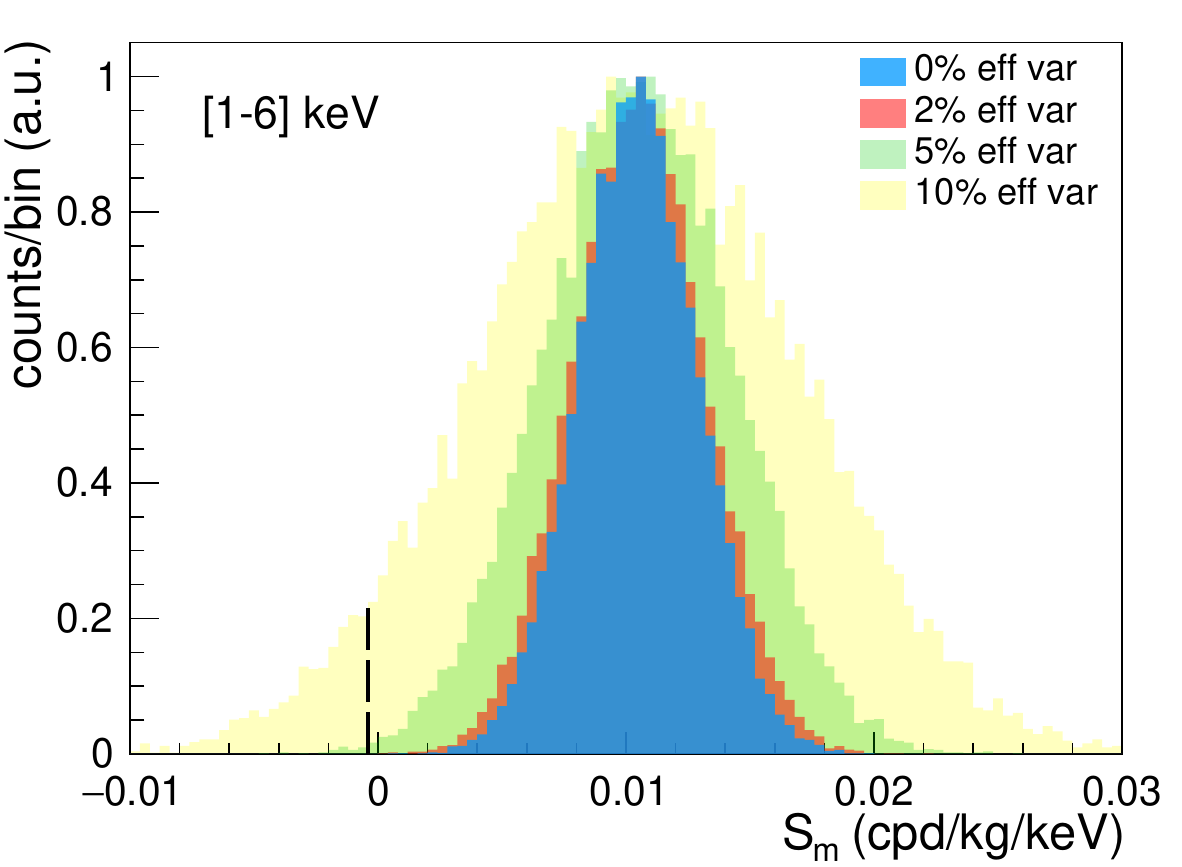}
    \end{minipage}
    \hfill  
    \begin{minipage}[b]{0.3\linewidth}
        \centering
        \includegraphics[width=\linewidth]{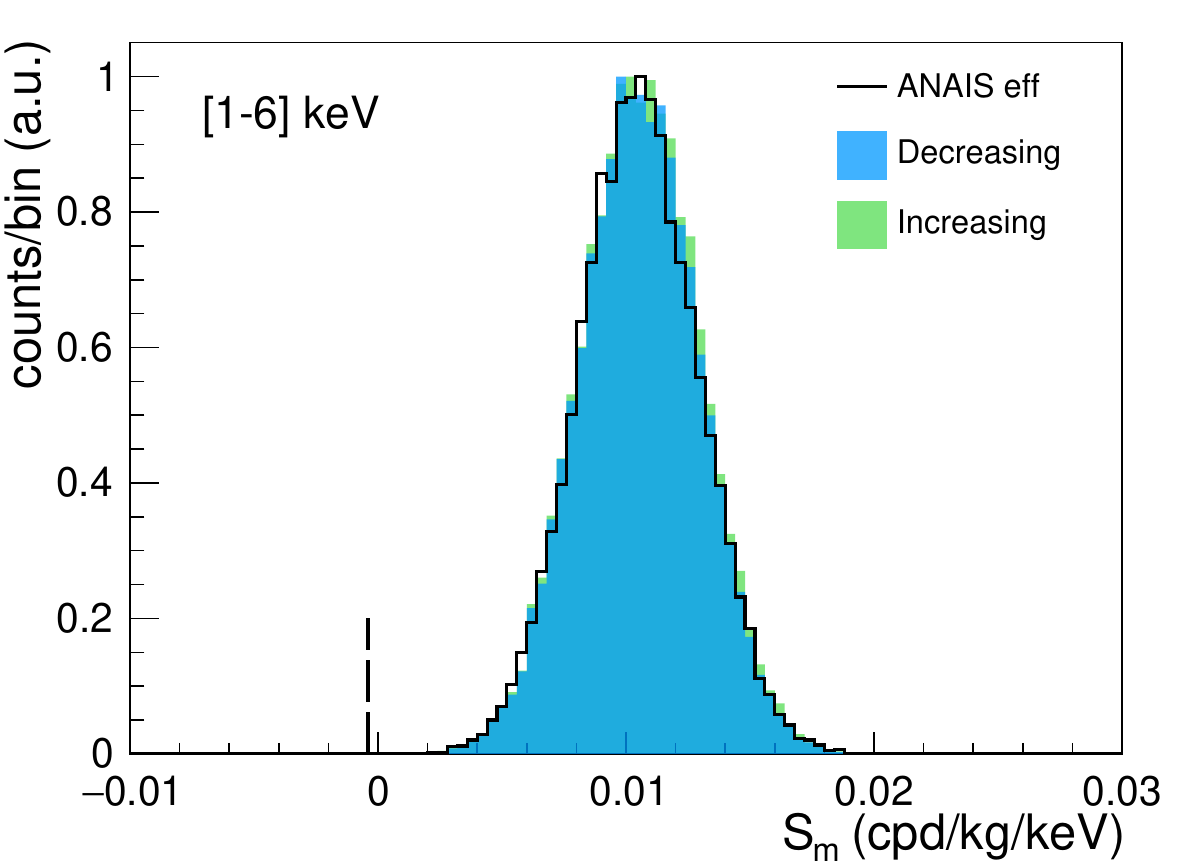}
    \end{minipage}
    \hfill
     \begin{minipage}[b]{0.3\linewidth}
        \centering
        \includegraphics[width=\linewidth]{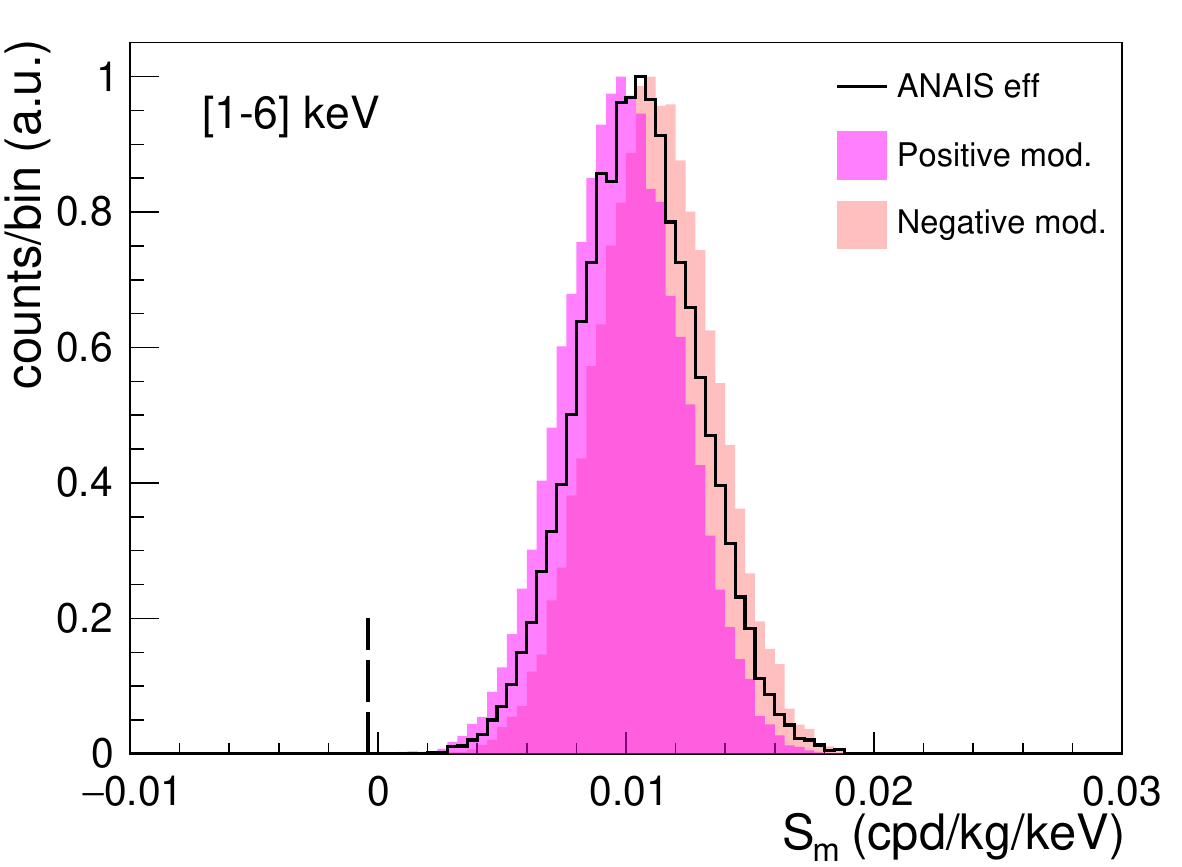}
    \end{minipage}
    \\[1ex]
    

        \begin{minipage}[b]{0.3\linewidth}
        \centering
        \includegraphics[width=\linewidth]{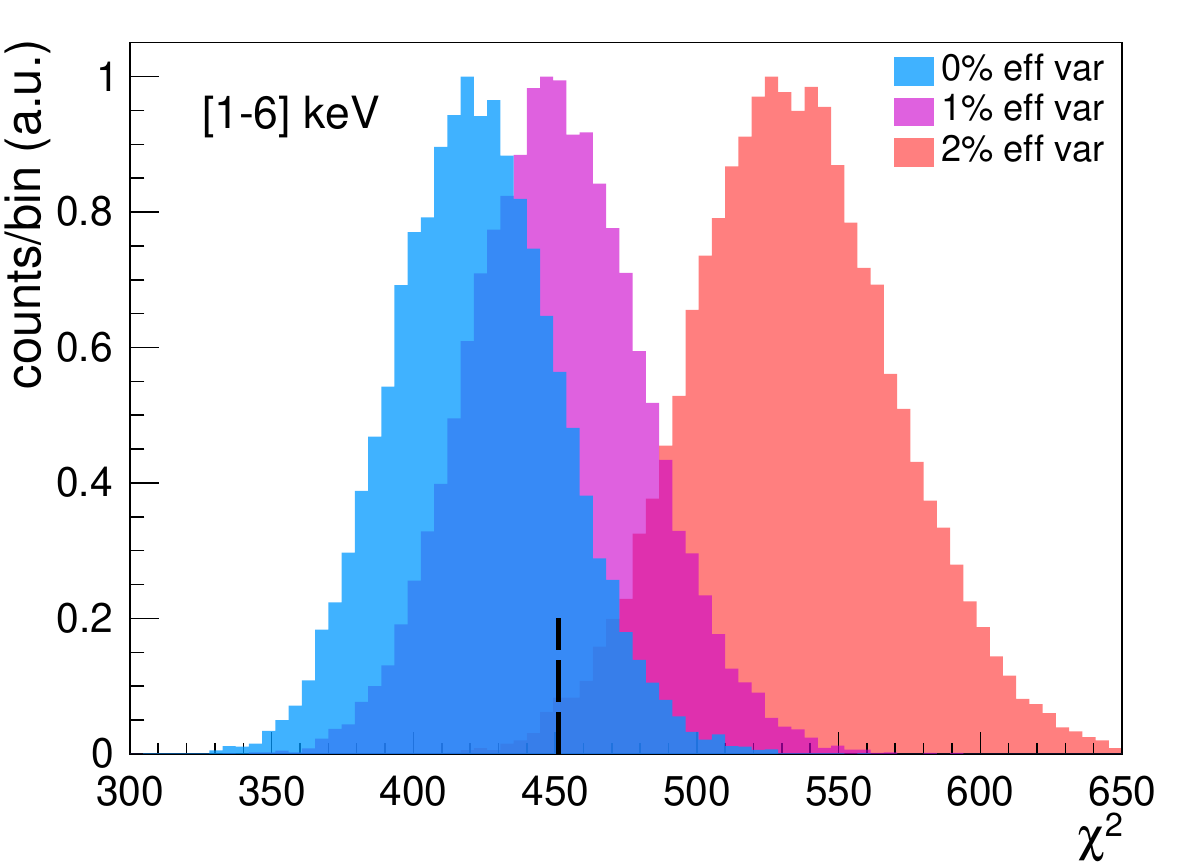}
    \end{minipage}   
    \hfill
    \begin{minipage}[b]{0.3\linewidth}
        \centering
        \includegraphics[width=\linewidth]{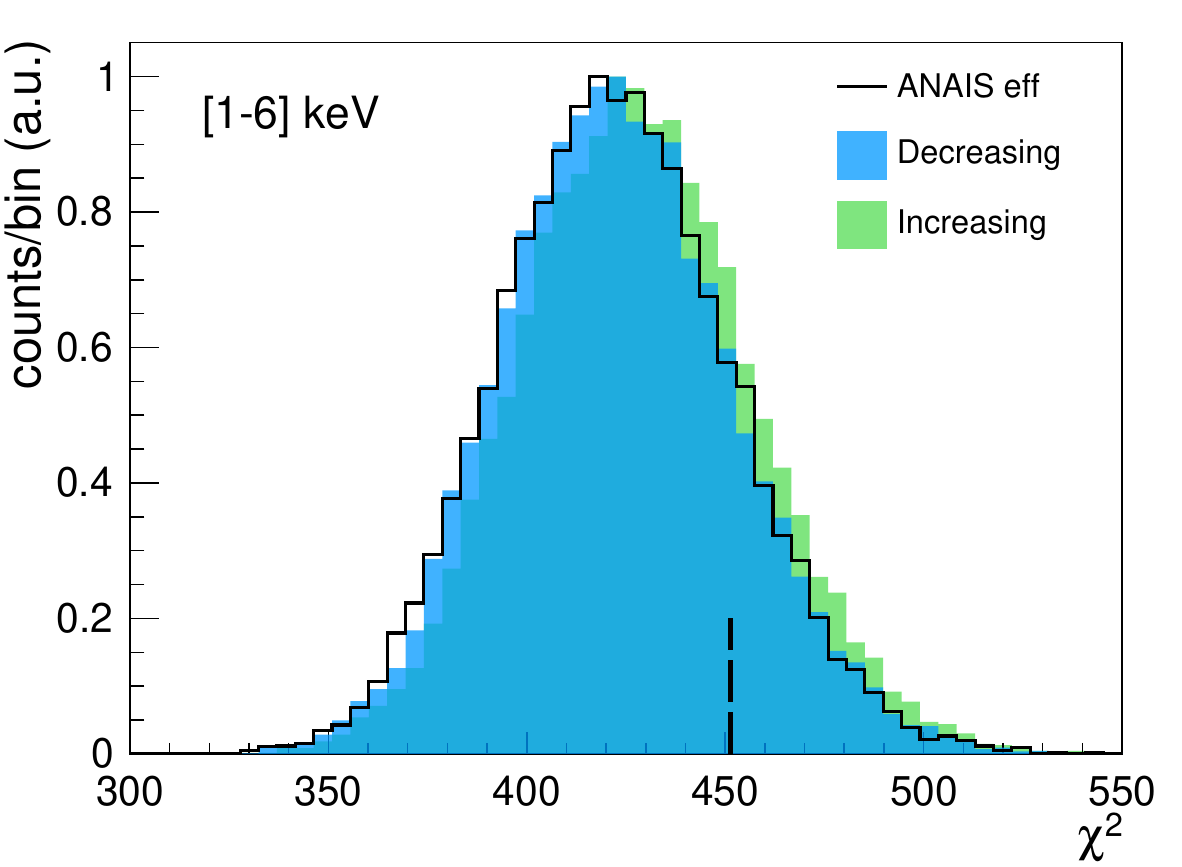}
    \end{minipage}
    \hfill
    \begin{minipage}[b]{0.3\linewidth}
        \centering
        \includegraphics[width=\linewidth]{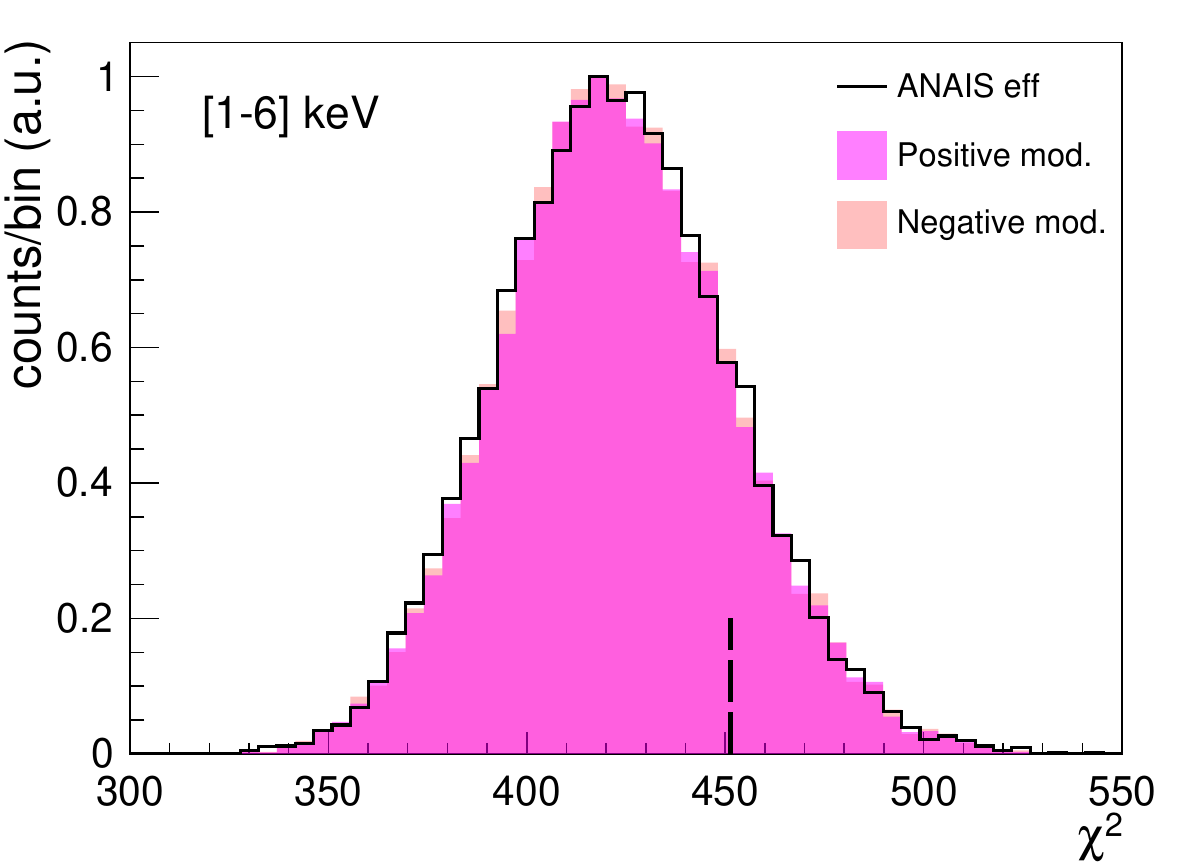}
    \end{minipage}

    \end{center}

    \caption{Results of 20\,000 toy MC simulations using the updated ANAIS--112 experimental features for 6 years, adding the modulation observed by DAMA/LIBRA. Upper panels: distribution of modulation amplitudes recovered in the [1--6]\,keV energy region for fluctuations in the efficiency of 0, 2, 5 and 10$\%$ (left panel); for efficiencies with a 0.6\% linear variation with time in all detectors, decreasing and increasing (middle panel) and for annually modulated (or antimodulated) efficiencies in 2 modules at 0.1\% and constant for the rest (right panel). Lower panels: corresponding $\chi^2$ value distribution of the fits (ndf=422). The \ANAIS result is shown as dashed line in all the panels. }
    \label{fig:depeff}
\end{figure*}

Then, we have introduced variations in the detection efficiencies of different magnitudes around the mean value and analyzed the modulation hypothesis. The modulation amplitudes obtained by fitting the simulated rate evolution show distributions centered in the \DL value, with a standard deviation that increases with the fluctuation in the efficiencies introduced, as expected. Figure~\ref{fig:depeff} (left-upper panel) shows the distribution of the modulation amplitudes derived from the fits in the [1--6]\,keV energy region for fluctuations in the efficiency of 0, 2, 5 and 10$\%$, while the \ANAIS result is shown as a dashed line. 
The measured efficiency variations in \ANAIS are on the order of 0.3\,$\%$, which implies a negligible probability of obtaining a null result when a modulation as large as that observed by \DL is present in the data. 
It is worth highlighting that variations as large as 5\,$\%$ are required to yield a non-negligible probability of hindering a true modulation in the data. On the other hand, the $\chi^2$ value of the fit is highly sensitive to the efficiency fluctuations, as shown in the left-lower panel of Figure~\ref{fig:depeff}.
The $\chi^2$ values obtained by ANAIS are strongly incompatible with fluctuations larger than 2\,$\%$.
\par

In addition, taking into account a possible dependence with time of our efficiencies, shown in Figure~3 within the Supplemental Material~\cite{supp}, we have evaluated the effect in the results of our analysis. First, we searched for different time dependences that could be hidden within the uncertainties of the efficiencies, detector by detector, both a linear dependence (decreasing or increasing with time) and annual modulation or antimodulation.
\par
We prepared 20\,000 toy MC simulations of experiments equivalent to 6~years of \ANAIS data, with and without adding the modulation observed by the \DL experiment, considering efficiencies with a linear dependence on time in the [1--6] keV energy region. The maximum variation observed in the detectors that show this trend (D2, D6-D8) is 0.6\% around the mean value and this is the value that has been considered in the toy MCs for all detectors.
The effect is completely negligible both in the modulation amplitude and in the chi-square distribution corresponding to the fits, as can be seen in the middle panels of Figure~\ref{fig:depeff}. 
 
\par
On the other hand, no consistent modulation in the efficiency of the ANAIS modules is observed when fitting to a constant or a linear variation over time plus an annual modulation. Seven out of nine modules provide results compatible with the null hypothesis. However, two modules (D4 and D7) result in the fitting with antimodulated efficiencies at 0.1\% level. Because a modulation or antimodulation in the efficiency would produce a bias in the determination of the modulation amplitude, we studied the effect with the toy MCs, as done before, including an annual modulation (antimodulation) in the efficiency of 0.1\% in 2 modules and constant efficiencies for the other modules. Results are shown in the right panels of Figure \ref{fig:depeff}. The corresponding bias in the modulation amplitude derived from this analysis amounts to about 6.3\% of the DAMA modulation amplitude (higher or lower for antimodulation and modulation, respectively). Therefore, an antimodulation of the efficiencies at this level is not sufficient to mask the DAMA/LIBRA signal.

\begin{figure}
     \centering
     \includegraphics[width=0.41\textwidth]{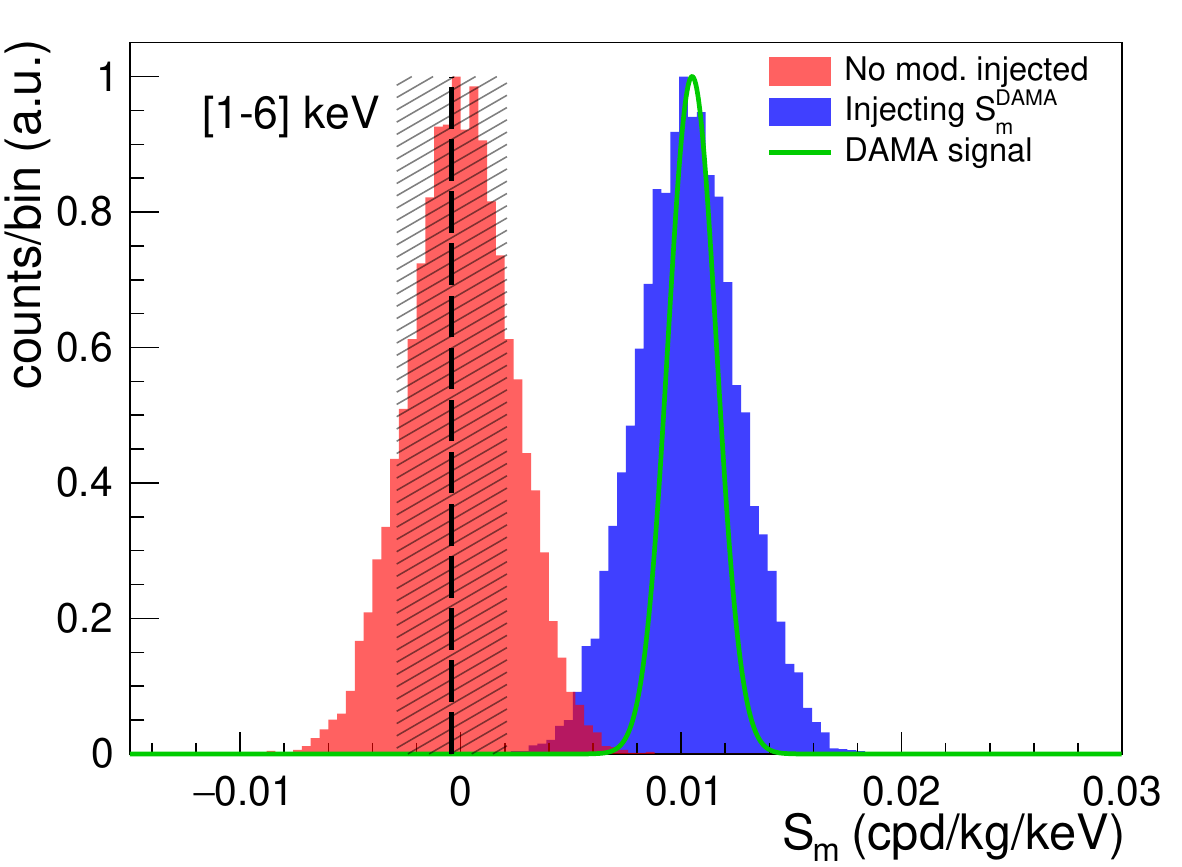} 

     \caption{Distribution of the modulation amplitudes recovered in the [1--6]~keV energy region with (blue) and without (red) injecting the \DL signal in 10\,000 toy MC simulations. The \ANAIS result is represented by the dashed black line, with the uncertainty shown as a pattern of black lines, and the \DL signal is displayed in green.}
   \label{fig:resMC}
\end{figure}

\par
Finally, we have also used equivalent toy MCs for evaluating the ANAIS--112 sensitivity to \DL result by analyzing the residuals found after subtracting the fitted background. These residuals are shown for our six-year data in Figure~8 within the Supplemental Material~\cite{supp}. Figure~\ref{fig:resMC} shows the distribution of the modulation amplitudes recovered with and without injecting the \DL modulation, which confirms the statistical significance of our result.

\subsection*{Model-independent comparison of results}

\begin{figure}
     \begin{minipage}[b]{0.9\linewidth}
     \centering 
     \includegraphics[width=\linewidth]{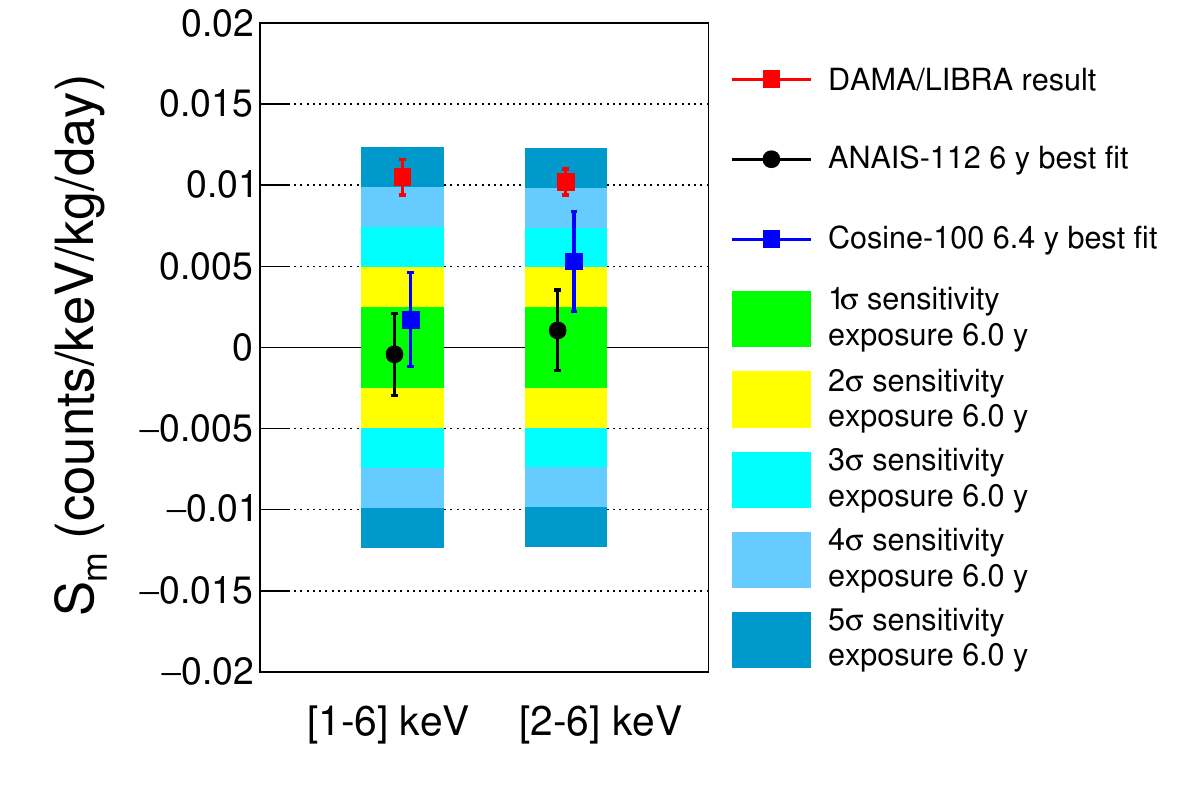} 
   \end{minipage}
   \hfill
    \caption{Comparison of the results from ANAIS--112, COSINE--100 \cite{Carlin:2024maf}, and \DL \cite{Bernabei:2020mon}.}
   \label{fig:comp}
   
\end{figure}

Figure~\ref{fig:comp} compares the results of ANAIS--112 (this work), COSINE--100~\cite{Carlin:2024maf} and \DL~\cite{Bernabei:2020mon}. The colored bands correspond to the ANAIS--112 sensitivity estimates for the six years exposure. It is worth to highlight that this comparison is independent from the DM particle and halo models and the main systematics remaining is that related with the scintillation QF for NR (already commented in the Letter). 

\clearpage
\setcounter{table}{0}
\setcounter{figure}{0}

\section*{Supplemental Material}

\subsection*{Detector stability}

\begin{figure*}
    \centering
    \includegraphics[width=0.75\textwidth]{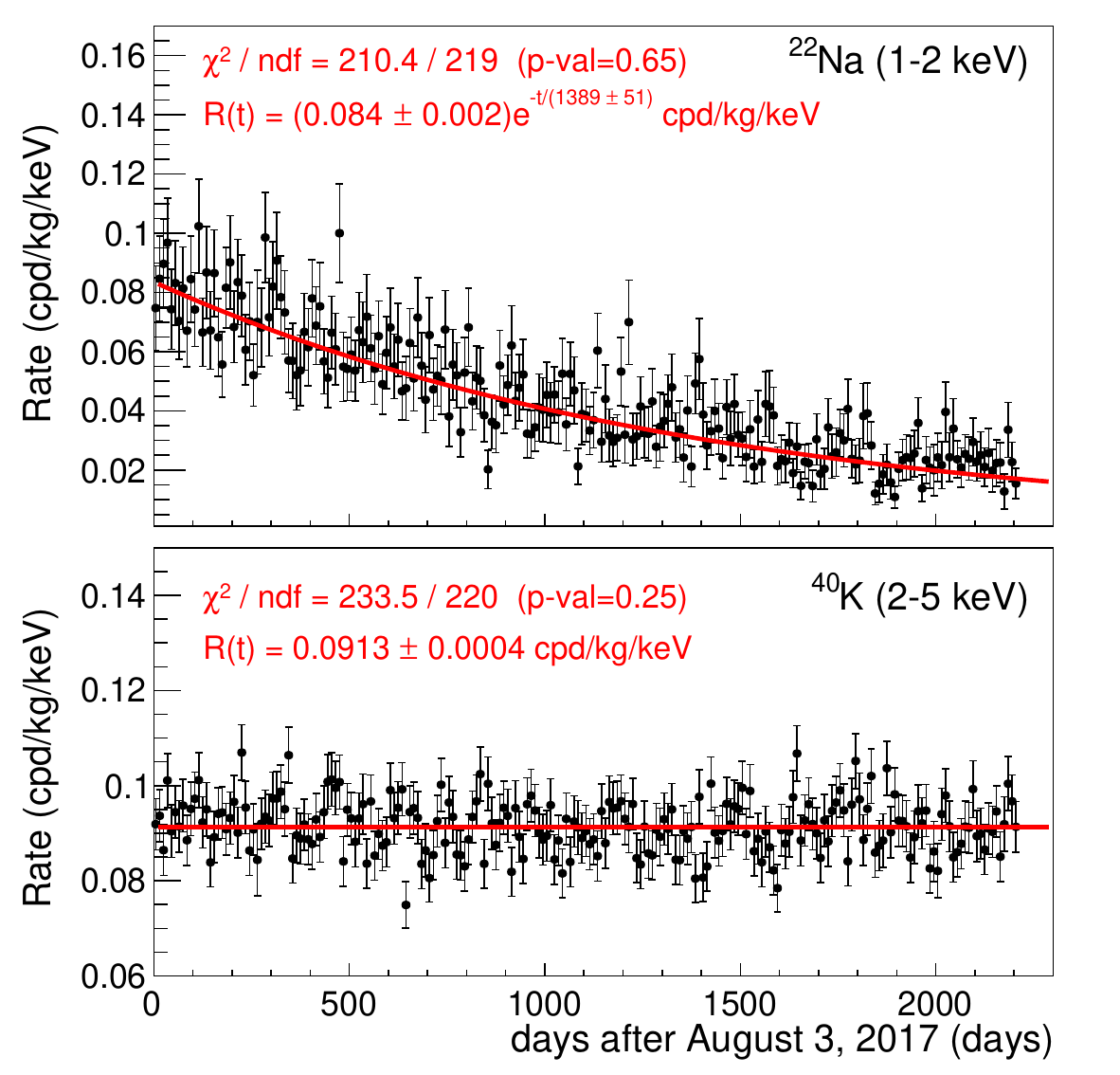}
     \caption{Time evolution of the rate of events corresponding to \Na (upper panel) and \K (lower panel) at low energy identified by the coincidence with the corresponding high-energy gamma in a second module. }
     \label{fig:NaKEvol}
\end{figure*}

\begin{figure*}
    \centering
    \includegraphics[width=\textwidth]{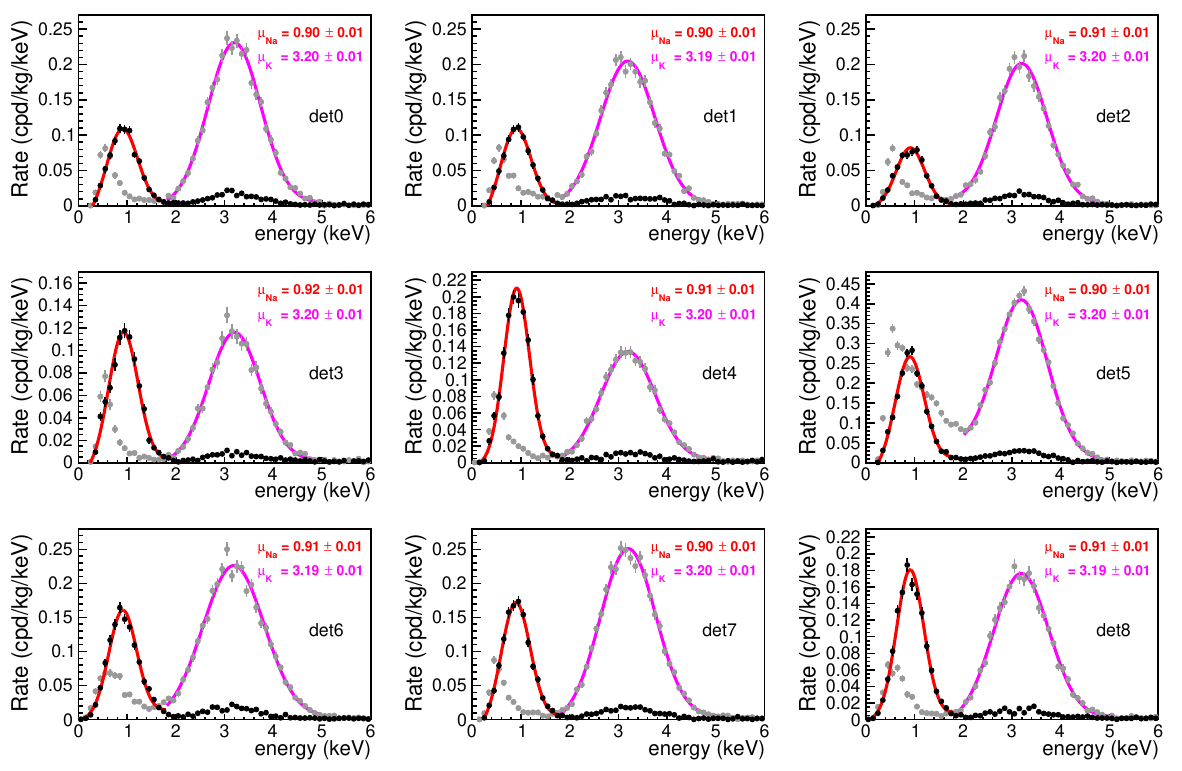}
     \caption{Low energy spectra of the nine modules in coincidence with a high-energy gamma in the range [1200–1340] keV (black points) and [1340–1560] keV (gray points) in a second module (after recalibration described in the introduction section). Data correspond to the 6-year exposure of ANAIS--112. Peaks at 0.9 and 3.2 keV, attributed to \Na and \K decay in the NaI bulk, respectively, can be clearly observed. Solid lines: fit to a Gaussian plus linear background lineshape. Mean value of the energy corresponding to the main peaks is also shown in each panel with the corresponding uncertainty, in keV.}
     \label{fig:NaKpeaks}
\end{figure*}

Many checks to guarantee a stable energy threshold along the 6 years of data taking have been carried out. Figure~\ref{fig:NaKEvol} shows the time evolution of the rate of events corresponding to \Na and \K at 0.9 and 3.2~keV, respectively, selected by the coincidence with a high-energy gamma in a second module. \Na events, in particular, fell below the 1~keV analysis threshold, and \K events are fully contained in the ROI of the ANAIS--112 experiment. Both rates (integrated for the nine modules) show the expected time behavior, a constant rate in the case of \K and an exponential decay compatible with the isotope lifetime in the case of \Na (1389$\pm$51~d is the lifetime derived from the fit, while 1369~d is the nominal lifetime for $^{22}$Na). Instabilities, either in the energy calibration or in the energy threshold, would have been identified by an anomalous time evolution of these rates.
Figure~\ref{fig:NaKpeaks} shows the \Na and \K events al low energy, normalized to cpd/kg/keV, selected by the coincidence with the corresponding high-energy gamma in a second module, and accumulated along the 6-year exposure of the experiment. These peaks are obtained by adding all the background runs conveniently corrected for gain drifts using the procedure explained in the introduction section and then, recalibrated. In case the gain correction was introducing additional uncertainties, these peaks would appear deformed and with larger resolution than expected. None anomalous effect has been observed. While \K events are evenly distributed along the 6-years of data taking, \Na events are scarce in the last year of the analysed exposure. It can be observed well below the 1~keV analysis threshold for \K events the peak corresponding to the L-shell-EC, confirming a sound trigger for events in the sub-keV energy range. 

\begin{figure*}
    \centering
    \includegraphics[width=\textwidth]{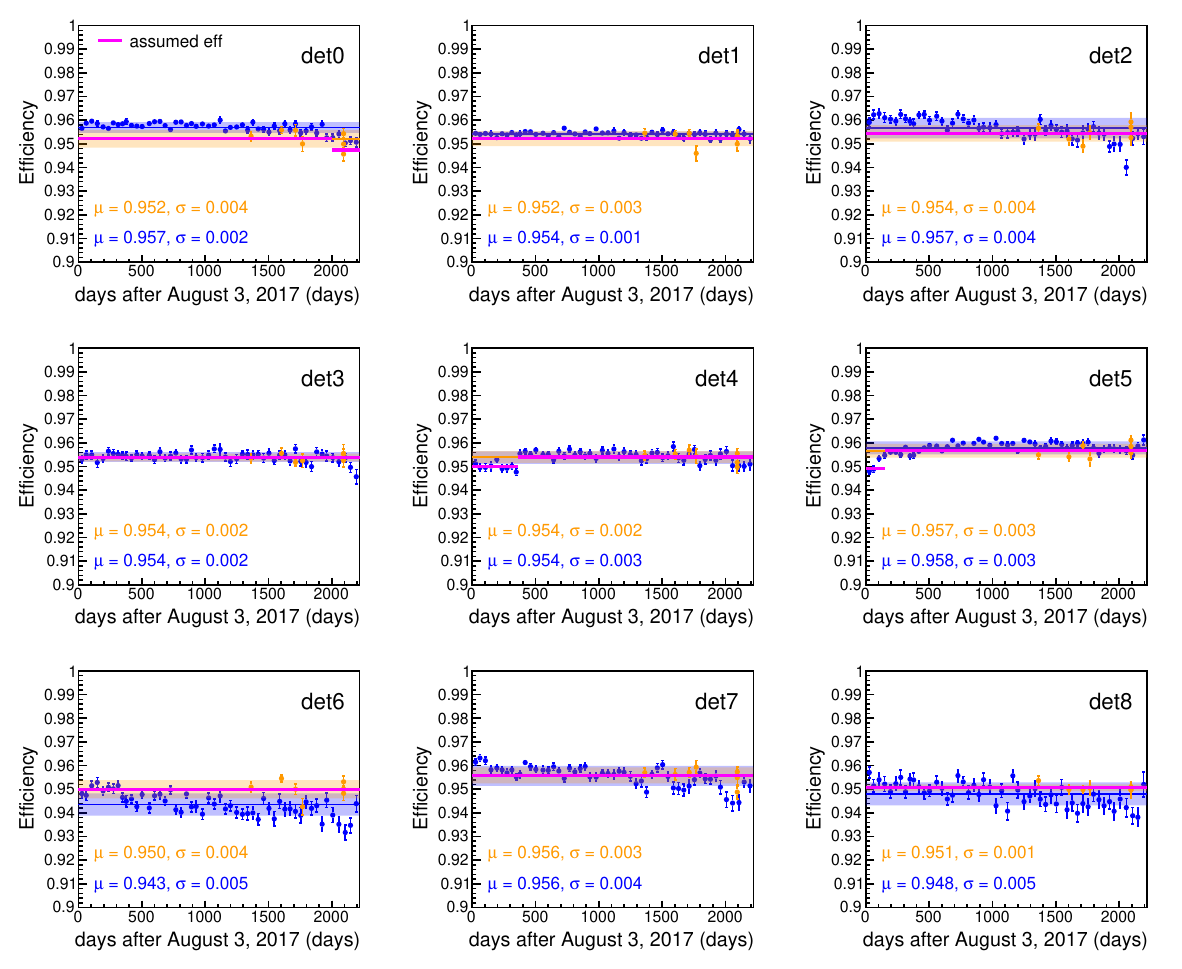}
    \caption{
     Evolution of the total detection efficiency in [1--6]\,keV estimated from \Cd calibration (in blue) and neutron calibration (in orange) along the six years of data taking for the nine \ANAIS modules. The solid blue and orange lines correspond to the mean values for each detector, and the shaded regions represent the standard deviations. The mean value and the standard deviation for each module are also shown in the panels. The efficiency used in the analysis is also represented by a solid magenta line.}
     \label{fig:effEvol}
\end{figure*}

Figure~\ref{fig:effEvol} shows the time evolution of the total detection efficiency in the ROI of \ANAIS ([1--6]\,keV), estimated using both, \Cd calibration events (in blue) and neutron calibration events (in orange). Neutron calibrations are not evenly distributed along the data taking, but are concentrated in the second half. The solid blue and orange lines correspond to the mean values for each detector, and the shaded regions represent the standard deviations. Both are compatible within uncertainties. \Cd data allow to consider constant efficiencies in all the modules, but D0, D4 and D5, as explained in the introduction section. The efficiency used in the analysis is also represented in Figure~\ref{fig:effEvol} by a solid magenta line.
  
\subsection*{Low energy background}
\begin{figure*}
    \centering
    \includegraphics[width=0.6\textwidth]{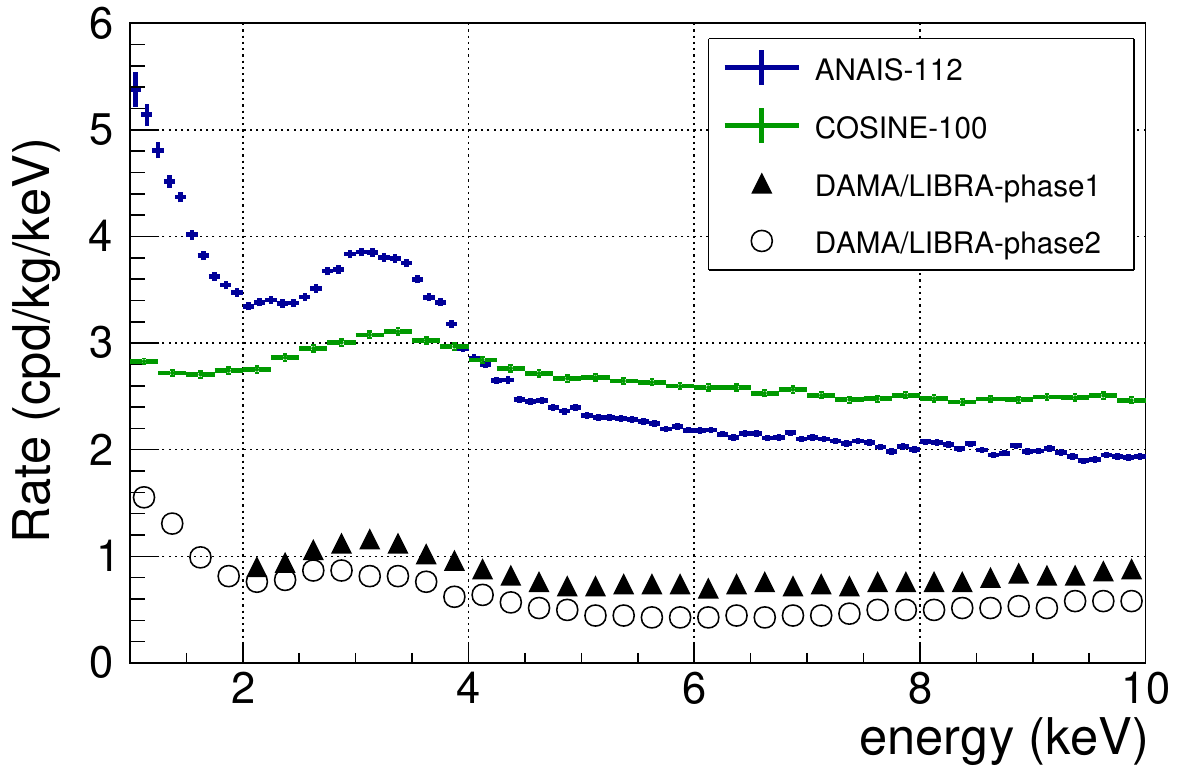}
    \caption{Comparison of the total single-hit energy spectrum measured after six years of data in \ANAIS (blue points), COSINE--100~\cite{COSINE-100:2021xqn} (green points) and DAMA/LIBRA~\cite{Bernabei:2020mon} (phase1, black triangles, and phase2, open circles).}
     \label{fig:bkg}
\end{figure*}

\ANAIS background in the ROI after applying the event selection procedures based on BDT and explained in the introduction section is shown in Figure~\ref{fig:bkg}. \COSINE and \DL backgrounds are also shown for comparison. It can be clearly noticed the \K peak at 3.2~keV present in \ANAIS data and the excess of events below 2~keV which is still under study and the background model is not able to account for. 
%

\subsection*{Exposure}
The accumulated exposure used for the annual modulation analysis corresponding to the six years of data of the \ANAIS experiment is summarized in Table~\ref{tab:exposure}. It also details the dead time (measured using latched counters during the data taking), down time (primarily due to bi-weekly $^{109}$Cd calibrations and seven $^{252}$Cf calibrations carried out in the referred period), 
percentage of live time rejected in the analysis, and the corresponding effective exposure after subtracting the latter. 
\par

\setlength{\tabcolsep}{0.8em}
\begin{table*}[htbp]
\centering
\begin{tabular}{ccccccc}
\hline\hline
Time period & Exposure & Dead time & Down time & \multicolumn{2}{c}{Rejected periods (\%)} & Effective exposure  \\
 & (kg$\times$yr) & (\%) & (\%) & muon cut & rate cut & (kg$\times$yr)\\
\hline
Aug 3, 2017 – July 31, 2018 & 104.80 & 2.88 & 3.19 & 2.64 & 0.60 & 101.19\\
Aug 1, 2018 – Aug 28, 2019  & 115.39 & 2.07 & 2.42 & 2.64 & 0.38 & 111.75\\
Aug 29, 2019 – Aug 13, 2020 & 102.86 & 2.38 & 2.54 & 2.53 & 0.38 &  99.72\\
Aug 14, 2020 – Aug 3, 2021  & 104.40 & 2.42 & 2.44 & 2.59 & 0.34 & 101.19\\
Aug 4, 2021 – Aug 30, 2022  & 116.86 & 1.89 & 1.39 & 2.83 & 0.28 & 113.12\\
Aug 31, 2022 – Aug 17, 2023 & 102.24 & 1.80 & 3.96 & 2.74 & 0.46 &  98.78\\
\hline
TOTAL & 646.55 & & & & & 625.75 \\
\hline \hline
\end{tabular}
\caption{\label{tab:exposure} 
For each of the six years of data collection, first column: start and end dates; second column: exposure calculated by multiplying live time by mass (112.5~kg); third column: percentage of dead time; fourth column: percentage of down time; fifth and sixth columns: percentages of time with respect to the live time corresponding to the two types of rejected periods (one second after a muon triggering the veto and one day if the daily averaged rate is above three standard deviations from the annually averaged rate, respectively); last column: effective exposure after subtracting the rejected periods.
}
\end{table*}

\subsection*{Annual modulation fit results}
\begin{figure*}
    \centering
    \includegraphics[width=0.85\textwidth]{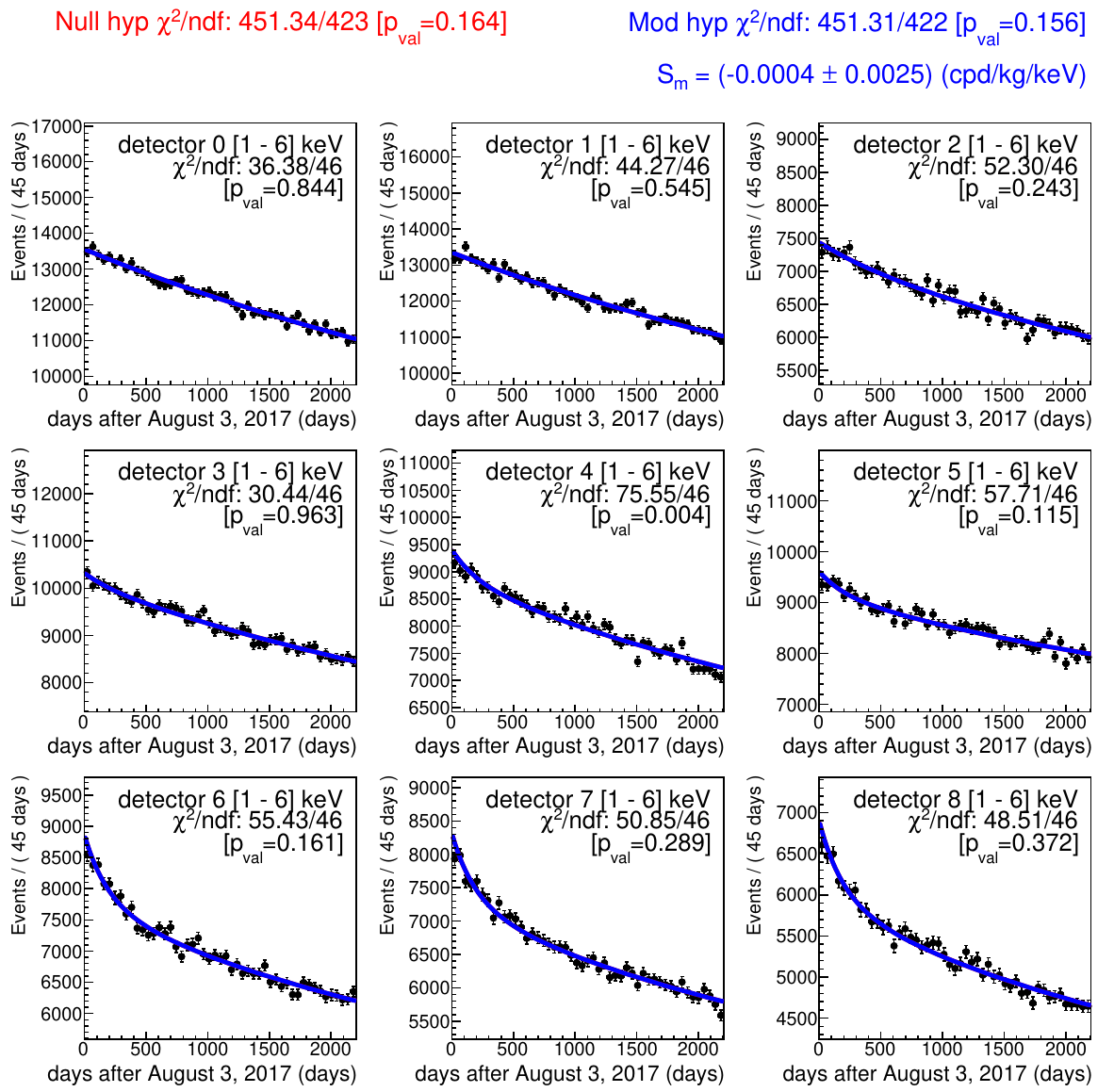}
    \caption{
    Results of the fit for the data from the nine modules in the [1--6]\,keV energy region, under the modulation (blue) and null hypotheses (red). 
    In all the panels, the red line is masked by the blue one, 
as the fit obtained for the modulated hypothesis is consistent with $S_m=0$. $\chi^2$/ndf and p-values under the modulation hypothesis are also individually displayed for each module.}
    \label{fig:rateEvol16supp}
\end{figure*}

\begin{figure*}
    \centering
    \includegraphics[width=0.85\textwidth]{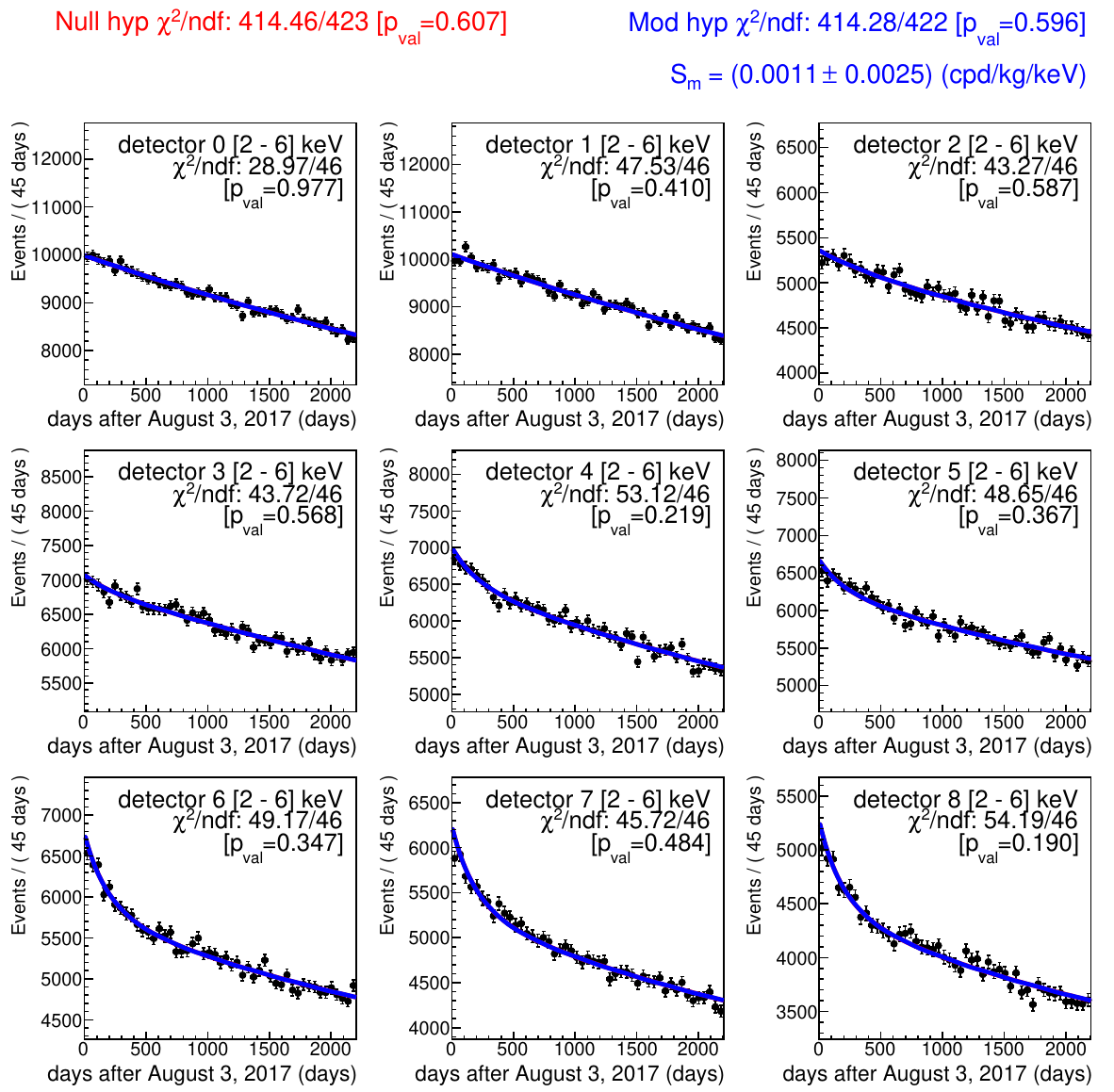}
    \caption{
    Results of the fit for the data from the nine modules in the [2--6]\,keV energy region, under the modulation (blue) and null hypotheses (red). 
    In all the panels, the red line is masked by the blue one, 
as the fit obtained for the modulated hypothesis is consistent with $S_m=0$. $\chi^2$/ndf and p-values under the modulation hypothesis are also individually displayed for each module.}
    \label{fig:rateEvol26}
\end{figure*}

\begin{figure*}
    \centering
    \includegraphics[width=0.85\textwidth]{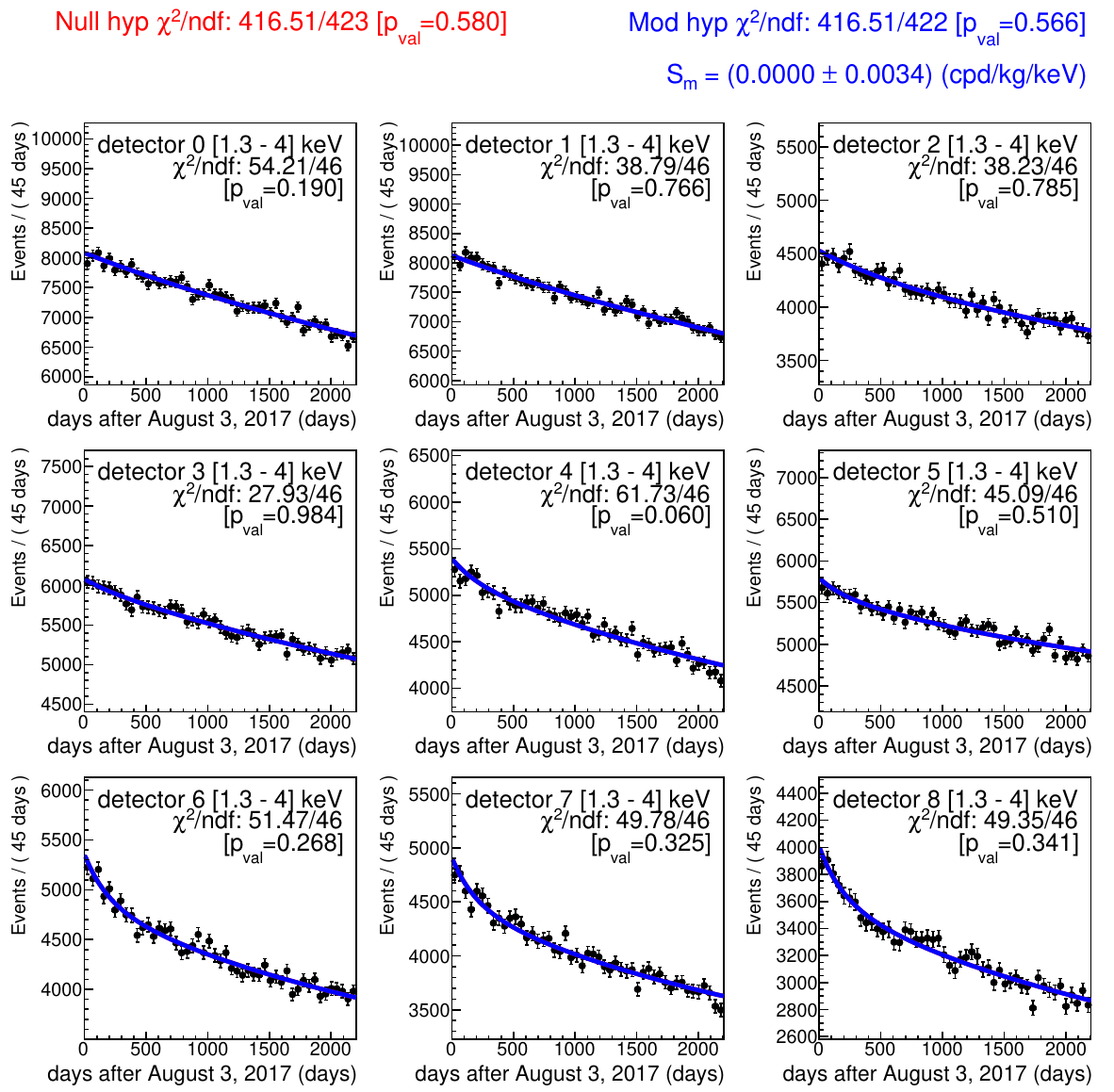}
    \caption{
    Results of the fit for the data from the nine modules in the [1.3--4]\,keV energy region, corresponding to the sodium nuclear recoil energy range of [6.7--20]~\keVnr, under the modulation (blue) and null hypotheses (red). 
    In all the panels, the red line is masked by the blue one, 
as the fit obtained for the modulated hypothesis is consistent with $S_m=0$. $\chi^2$/ndf and p-values under the modulation hypothesis are also individually displayed for each module.}
    \label{fig:rateEvol14}
\end{figure*}

\begin{figure*}
    \centering
    \includegraphics[width=\textwidth]{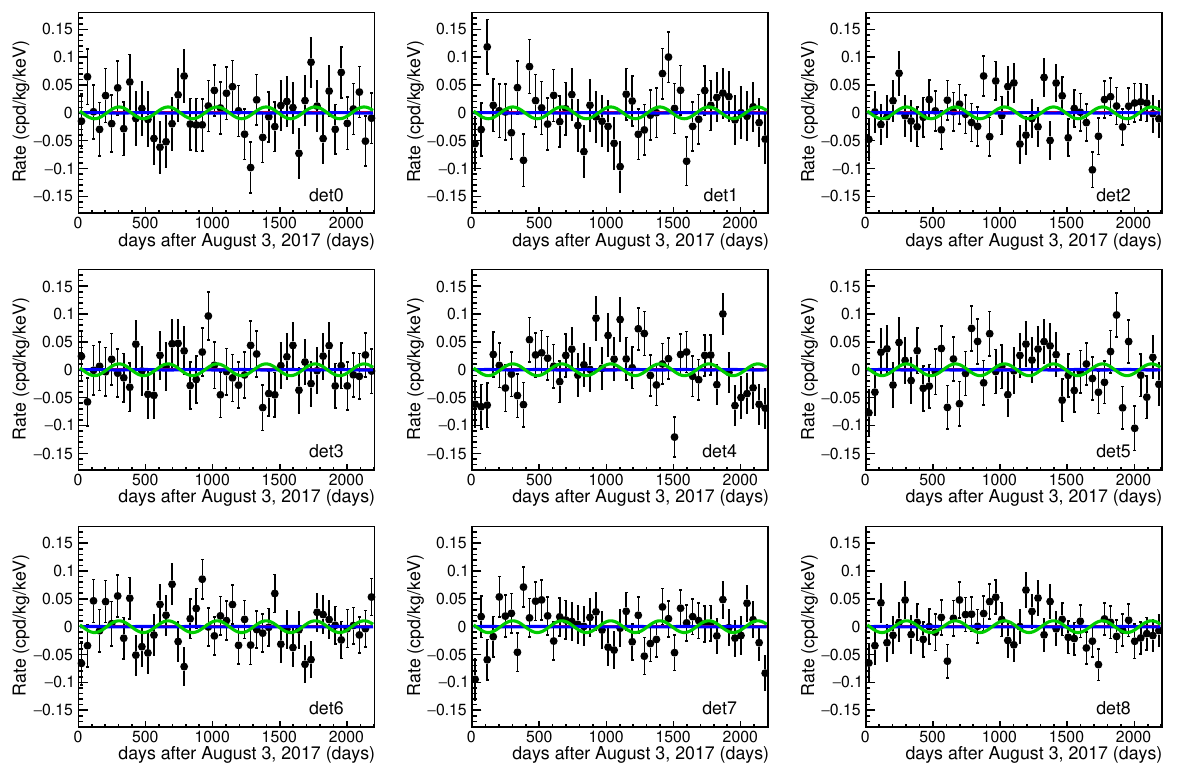}
    \caption{
    Fit results for the data from the nine modules in the [1--6]\,keV energy region after subtracting the non-modulated term from Eq.~2. Blue and red lines are the result of the fit for the modulation and null hypothesis, respectively, after subtracting the non-modulated term from Eq.~2. The modulation observed by \DL is shown in green.}
     \label{fig:residualsEvol16Det}
\end{figure*}

The results of the fits for the annual modulation analysis in the three energy regions considered for the nine modules are shown in Figures~\ref{fig:rateEvol16supp}, \ref{fig:rateEvol26} and \ref{fig:rateEvol14}. $\chi^2$/ndf and p-values under the modulation hypothesis are also individually displayed for each module, besides the global analysis $\chi^2$/ndf and  p-values for the null and modulation hypothesis and the $S_m$ corresponding to the best fit for the latter. In all the energy regions and detectors, we obtain good fits to the PDF built with our MC background model time evolution. We can conclude that our data are fully consistent with no modulation and incompatible with DAMA/LIBRA. Fit results for the modulation amplitude and corresponding p-values were shown in Table-I in the Letter.  Table~\ref{tab:nuisance} reports on the nuisance parameters derived from the fits, both background index and $f$ for each module, which measures the deviation of the fit from the MC background model detector by detector. Only detector 5 shows systematically values of $f$ below 85\%. Figure~\ref{fig:residualsEvol16Det} shows the results of the fit after subtracting the non-modulated component. The modulation observed by DAMA/LIBRA is shown in the same plot for comparison. 

%

%
\begin{table*}
\centering
\begin{tabular}{ccccccc}
\hline\hline

\multirow{3}{*}{Detector} & \multicolumn{2}{c}{[1--6] keV} & \multicolumn{2}{c}{[2--6] keV} & \multicolumn{2}{c}{[6.7--20] \keVnr} \\
                          & Bkg index & \(f\) & Bkg index & \(f\) & Bkg index & \(f\) \\ 
                          & (cpd/kg/keV) & & (cpd/kg/keV) & & (cpd/kg/3.3 \keVnr) & \\ \hline
0                         & 4.711$\pm$0.007 & 0.98$\pm$0.02 & 4.396$\pm$0.007 & 0.97$\pm$0.03 & 3.512$\pm$0.006 & 0.98$\pm$0.03 \\ \hline
1                         & 4.672$\pm$0.007 & 0.90$\pm$0.02 & 4.438$\pm$0.007 & 0.98$\pm$0.03 & 3.550$\pm$0.006 & 0.91$\pm$0.03 \\ \hline
2                         & 2.550$\pm$0.005 & 0.96$\pm$0.03 & 2.334$\pm$0.005 & 0.97$\pm$0.04 & 1.959$\pm$0.004 & 0.90$\pm$0.04 \\ \hline
3                         & 3.563$\pm$0.006 & 0.85$\pm$0.03 & 3.064$\pm$0.006 & 0.94$\pm$0.03 & 2.637$\pm$0.005 & 0.85$\pm$0.03 \\ \hline
4                         & 3.100$\pm$0.005 & 0.88$\pm$0.02 & 2.871$\pm$0.005 & 0.95$\pm$0.03 & 2.244$\pm$0.005 & 0.86$\pm$0.03 \\ \hline
5                         & 3.308$\pm$0.06  & 0.63$\pm$0.02 & 2.805$\pm$0.005 & 0.80$\pm$0.03 & 2.507$\pm$0.005 & 0.64$\pm$0.03 \\ \hline
6                         & 2.698$\pm$0.005 & 0.95$\pm$0.02 & 2.572$\pm$0.005 & 0.99$\pm$0.02 & 2.097$\pm$0.005 & 0.99$\pm$0.03 \\ \hline
7                         & 2.525$\pm$0.005 & 0.96$\pm$0.02 & 2.337$\pm$0.005 & 1.05$\pm$0.03 & 1.934$\pm$0.004 & 0.96$\pm$0.03 \\ \hline
8                         & 2.049$\pm$0.004 & 0.97$\pm$0.02 & 1.956$\pm$0.004 & 0.99$\pm$0.03 & 1.546$\pm$0.004 & 0.98$\pm$0.03 \\ \hline
 
\hline \hline
\end{tabular} 
\caption{Summary of the nuisance parameters obtained in the fits searching for an annual modulation in the six years of \ANAIS data taking for different energy regions.}
\label{tab:nuisance}
\end{table*}

\subsection*{ANAIS Sensitivity prospects}
\begin{figure*}
     \centering
     \includegraphics[width=0.6\textwidth]{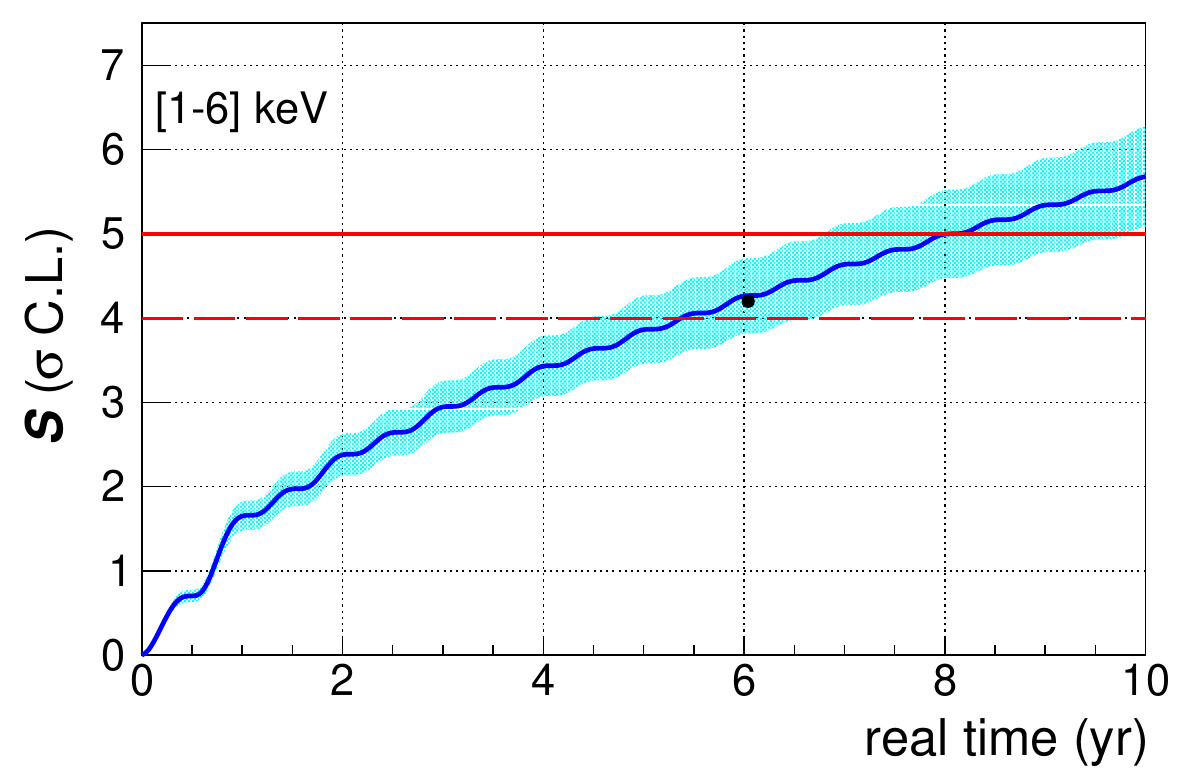} 

     \caption{\ANAIS sensitivity to the \DL signal in $\sigma$ C.L. units as a function of real time in the [1--6]\,keV energy region. The black dot is the sensitivity measured experimentally for 6-year exposure. The cyan band represents the 68\% C.L. \DL uncertainty.}
   \label{fig:sens}
\end{figure*}

Figure~\ref{fig:sens} displays in dark blue line the \ANAIS sensitivity projection in the [1--6]\,keV energy region following Ref.~\cite{Coarasa:2018qzs}, conveniently updated to the effective exposure, background level and detection efficiency presented in this work. Similar sensitivities are obtained for the [2--6]\,keV energy region. Cyan band takes into account the 68\% uncertainty in $S_m^{\text{DAMA}}$. The black dot is the sensitivity derived from the 6-year result presented here, in good concordance with our estimate. This result supports our expectation of achieving a 5\,$\sigma$ sensitivity to the \DL result by the end of 2025. 

\end{document}